\documentclass[twocolumn,twocolappendix]{aastex63}
\usepackage{ragged2e}
\usepackage{natbib}

\accepted{ApJ}

\shorttitle{Star Formation Activity in Nearby Quasars Hosts}
\shortauthors{Molina, J. et al.}

\begin{document}

\title{Enhanced Star Formation Efficiency in the Central Regions of Nearby Quasars Hosts}

\correspondingauthor{Ran Wang}
\email{rwangkiaa@pku.edu.cn}

\author[0000-0002-8136-8127]{Juan Molina}
\affil{Kavli Institute for Astronomy and Astrophysics, Peking University, Beijing 100871, China}

\author[0000-0001-6947-5846]{Luis C. Ho}
\affil{Kavli Institute for Astronomy and Astrophysics, Peking University, Beijing 100871, China}
\affiliation{Department of Astronomy, School of Physics, Peking University, Beijing 100871, China}

\author[0000-0003-4956-5742]{Ran Wang}
\affil{Kavli Institute for Astronomy and Astrophysics, Peking University, Beijing 100871, China}
\affiliation{Department of Astronomy, School of Physics, Peking University, Beijing 100871, China}

\author[0000-0002-4569-9009]{Jinyi Shangguan}
\affil{Max-Planck-Institut f\"{u}r Extraterrestrische Physik (MPE), Giessenbachstr., D-85748 Garching, Germany}

\author[0000-0002-8686-8737]{Franz E. Bauer}
\affil{Instituto de Astrof{\'{\i}}sica and Centro de Astroingenier{\'{\i}}a, Facultad de F{\'{i}}sica, Pontificia Universidad Cat{\'{o}}lica de Chile, Casilla 306, Santiago 22, Chile}
\affiliation{Millennium Institute of Astrophysics (MAS), Nuncio Monse{\~{n}}or S{\'{o}}tero Sanz 100, Providencia, Santiago, Chile} \affiliation{Space Science Institute, 4750 Walnut Street, Suite 205, Boulder, Colorado 80301}

\author[0000-0001-7568-6412]{Ezequiel Treister}
\affil{Instituto de Astrof{\'{\i}}sica and Centro de Astroingenier{\'{\i}}a, Facultad de F{\'{i}}sica, Pontificia Universidad Cat{\'{o}}lica de Chile, Casilla 306, Santiago 22, Chile}

\begin{abstract}
We combine Atacama Large Millimeter/submillimeter Array and Multi Unit Spectroscopic Explorer observations tracing the molecular gas, millimeter continuum, and ionized gas emission in six low-redshift ($z \lesssim 0.06$) Palomar-Green quasar host galaxies to investigate their ongoing star formation at $\sim$\,kpc-scale resolution. The AGN contribution to the cold dust emission and the optical emission-line flux is carefully removed to derive spatial distributions of the star formation rate (SFR), which, complemented with the molecular gas data, enables the mapping of the depletion time ($t_{\rm dep}$). We report ubiquitous star formation activity within the quasar host galaxies, with the majority of the ongoing star formation occurring in the galaxy center. The rise of the star formation rate surface density ($\Sigma_{\rm SFR}$) toward the nucleus is steeper than that observed for the cold molecular gas surface density, reaching values up to $\Sigma_{\rm SFR} \approx 0.15-0.80$\,$M_\odot$\,yr$^{-1}$\,kpc$^{-2}$. The gas in the nuclear regions is converted into stars at a shortened depletion time ($t_{\rm dep} \approx 0.2-2.0\,$Gyr), suggesting that those zones can be deemed as starbursts. At large galactocentric radius, we find that the ongoing star formation takes place within spiral arms or H\,{\sc ii} region complexes, with an efficiency comparable to that reported for nearby inactive spirals ($t_{\rm dep} \approx 1.8\,$Gyr). We find no evidence of star formation activity shutoff in the PG quasar host galaxies. On the contrary, these observations shed light on how the central environments of galaxies hosting actively accreting supermassive black holes builds up stellar mass.
\end{abstract}

\section{Introduction}

The correlations between the mass of supermassive black holes (BHs) and the host galaxy bulge properties \citep{Magorrian1998,Ferrarese2000,Gebhardt2000} have been commonly interpreted as evidence that BHs and galaxies coevolve by regulating each other's growth \citep{Kormendy2013}, and feedback from active galactic nuclei (AGNs) is widely accepted as the likely coupling mechanism \citep{Silk1998,Heckman2014,Harrison2018}. The link between BH accretion and star formation activity arises naturally because both phenomena are fed by the same gas reservoir and governed largely by similar secular processes that drive gas inward \citep{Kormendy2004}. The peak of BH accretion likely occurs a few hundred Myrs after the onset of starburst activity \citep{Wild2010}. During the accretion of mass onto the BH, an enormous amount of energy is released to the surroundings. Coupling merely a few percent of this output energy with the interstellar medium can heat and/or blow away the gas from the host galaxy (e.g., \citealt{Croton2006,Somerville2008,Schaye2015,Sijacki2015}). The growth of the BH is self-limiting, and it may simultaneously quench ongoing star formation activity within the host galaxy \citep{Fabian2012,Dubois2016}. 

Whether AGN feedback mediates BH-galaxy coevolution is still uncertain. Reports of suppressed star formation activity in local low-luminosity \citep{Ho2003,Ellison2016,Leslie2016,Jackson2020} and luminous \citep{Scholtz2018,Stemo2020} AGNs have upheld the notion that AGN feedback quenches star formation, driving the evolution of galaxies from star-forming to passive systems. Conversely, local Seyferts often show ongoing or recent star formation (e.g., \citealt{Davies2007,Esquej2014}), and compelling evidence indicates that the more luminous AGNs are commonly hosted in galaxies with star formation rates (SFRs) similar to those reported in typical inactive spirals \citep{Harrison2012,Rosario2012,Rosario2013,Husemann2014,Zhang2016,Stanley2017,Bernhard2019,Schulze2019,Grimmett2020,Koss2021,Vietri2022}, or even comparable to that of starbursts \citep{Young2014,Bernhard2016,Pitchford2016,Kirkpatrick2020,Shangguan2020b,Xie2021}. The increasing availability of cold gas observations targeting the carbon monoxide (CO) and neutral atomic hydrogen (H\,{\sc i}) emission lines (e.g., \citealt{Evans2001,Evans2006,Scoville2003,Ho2008,Xia2012,Brusa2015,Husemann2017,Kakkad2017,Husemann2019b,Shangguan2020,Salvestrini2022}) or measurements from indirect tracers based on dust absorption and emission (e.g., \citealt{Lutz2018,Shangguan2018,Shangguan2019,Yesuf2019,Yesuf2020a,Yesuf2020b,Zhuang2020}) support the view that AGNs are preferentially observed in gas-rich and highly star-forming systems \citep{Florez2020,Jarvis2020,Koss2021}. Moreover, the combination of the gas masses and SFRs has enabled the quantification of the star formation efficiency (SFE\,$\equiv$\,SFR$/M_{\rm gas}$) or gas depletion time ($t_{\rm dep} \equiv \,$1/{\rm SFE}), offering a complementary probe of how the cold gas is converted into stars in AGN host galaxies \citep{Husemann2017,Jarvis2020,Shangguan2020b,Koss2021,Zhuang2021}.

Another pathway to further study the interplay between AGN feedback and star formation is to spatially resolve the host galaxy spectroscopically (e.g., \citealt{Jahnke2004,Lipari2009,Husemann2014,Harrison2016,Husemann2017,Ilha2019,Feruglio2020,Kakkad2020,Lacerda2020,Riffel2021,Scholtz2021}). In integral-field spectroscopy (IFU) observations, the AGN emission is seen as blurred component following the observation point-spread function (PSF; e.g., \citealt{Husemann2016}), the essential constraint that allows to design an AGN-galaxy host deblending procedure so that the underlying host galaxy emission can be isolated and further analyzed \citep{Husemann2013,Rupke2017}. However, in these observations, the AGN flux is concentrated within a PSF-sized region, implying that recovering the host galaxy emission near this zone is devious, and extrapolation methods have to be adopted to minimize oversubtraction of the host galaxy emission (e.g., \citealt{Husemann2022}). On the other hand, observations resolving the molecular gas and/or cold dust content in the sub-millimeter/millimeter (sub-mm/mm) provide a cleaner view of the host galaxies (e.g., \citealt{Molina2021,Girdhar2022,RamosAlmeida2022}), and their combination with the IFU data offers a more complete picture of the conditions and dynamics of the interstellar medium around AGNs \citep{Husemann2019a,Rosario2019,Shimizu2019,Feruglio2020,Lamperti2021}.

\begin{table*}
	\centering
	\def\arraystretch{1.2}
	\setlength\tabcolsep{3pt}
    	\caption{\label{tab:sample} Basic Parameters of the Sample}
    	\vspace{0.2mm}
	\begin{tabular}{cccccccccccc}
		\hline
		\hline
		Object & R.A. & Decl. & $z$ & $D_L$ & Morphology & $\log M_\star$ & $\log M_{\rm H_2}$ & SFR$_{\rm IR}$ & $\log M_{\rm BH}$ & $\log L_{\rm 5100}$ & $L_{\rm bol}/L_{\rm Edd}$ \\
		& (J2000.0) & (J2000.0) & & (Mpc) & &($M_\odot$) & ($M_\odot$) & ($M_\odot$\,yr$^{-1}$) & ($M_\odot$) & (erg\,s$^{-1}$) & \\
		(1) & (2) & (3) & (4) &(5) & (6) & (7) & (8)& (9) & (10) & (11) & (12)\\
		\hline
	PG\,0050+124 & 00:53:34.94 & +12:41:36.2 & 0.061 & 282.3 & Disk & 11.12 & 10.2 & 26.3 & 7.57 & 44.76 &1.2\\
         PG\,0923+129 & 09:26:03.29 & +12:44:03.6 & 0.029 & 131.2 & Disk & 10.71 & 9.2 & 3.4 & 7.52 & 43.83 &0.2\\
         PG\,1011$-$040 & 10:14:20.69 & $-$04:18:40.5 & 0.058 & 267.9 & Disk & 10.87 & 9.5 & 2.9 & 7.43 & 44.23 &0.5\\
         PG\,1126$-$041 & 11:29:16.66 & $-$04:24:07.6 & 0.060 & 277.5 & Disk & 10.85 & 9.6 & 8.7 & 7.87 & 44.36 &0.2\\
        PG\,1244+026 & 12:46:35.25  & +02:22:08.8 & 0.048 & 220.1 & Disk & 10.19  & 8.9 & 2.1 & 6.62 & 43.77 &1.1\\
         PG\,2130+099 & 21:32:27.81 & +10:08:19.5 & 0.061 & 292.3 & Disk & 10.85 & 9.5 & 7.1 & 8.04 & 44.54 &0.3\\
		\hline
	\end{tabular}
	\justify
	{\justify \textsc{Note}--- (1) Source name. (2) Right ascension. (3) Declination. (4) Redshift. (5) Luminosity distance. (6) Morphology type of the host galaxy \citep{Zhang2016,Kim2017,Zhao2021} (7) Stellar mass; the $1\,\sigma$ uncertainty is 0.3\,dex \citep{Shangguan2018}. (8) Molecular gas mass inferred from CO measurements; the $1\,\sigma$ uncertainty is 0.3\,dex \citep{Shangguan2020}. (9) IR SED-based SFR estimated by adopting Eq.~4 of \citet{Kennicutt1998b} and a \cite{Kroupa2001} initial mass function; the typical uncertainty is $\sim 0.2-0.3$\,dex. (10) Black hole mass, estimated by applying the calibration of \citet{Ho2015} and taken from \citet{Shangguan2018}; $1\,\sigma$ uncertainty is 0.3\,dex. (11) AGN monochromatic luminosity at 5100\,\r{A}. (12) Eddington ratio, where $L_{\rm bol} = 10\, L_{\rm 5100}$ \citep{Richards2006} is the bolometric luminosity, and $L_{\rm Edd} = 1.26 \times 10^{38} (M_{\rm BH}/M_\odot)$\,erg\,s$^{-1}$ is the Eddington luminosity.}
\end{table*}

Within the active galaxy population, quasars---the most luminous AGNs---are the best-suited targets to explore the possible effect of AGN feedback on the ongoing star formation activity of the host galaxy. In the popular evolutionary scenario of \cite{Sanders1988}, a quasar is thought to be the product of the merging of two gas-rich systems, where the gas driven into the nuclear zone by gravitational torques fuels intense starburst activity and BH growth. The gas and dust enshrouding the nucleus are expelled by the overwhelming release of energy during the AGN phase, giving birth to an optically visible and largely unobscured quasar \citep{Hopkins2008,Treister2010}. The presence of young stellar populations is frequently associated with quasar host galaxies, in qualitative agreement with this evolutionary scenario, giving support to the notion that star formation accompanies or precedes BH growth \citep{Canalizo2001,Jahnke2007,Canalizo2013,Kim2019,DahmerHahn2022}. 

This work uses Multi Unit Spectroscopic Explorer (MUSE) IFU observations taken with the Very Large Telescope, in combination with Atacama Large Millimeter/submillimeter Array (ALMA) data, to measure the SFRs and molecular gas properties at $\sim$kpc-scales in six Palomar-Green (PG) quasar host galaxies \citep{Boroson1992}. We compute SFRs from H$\alpha$ and mm continuum fluxes decontaminated for additional emission sources, while the carbon monoxide (CO) emission line is used to estimate the molecular gas content. We conclude that, at least in low-redshift quasars, the increase of the SFR toward the nuclear zone of the host galaxy is steeper than that observed for the molecular gas component, implying that, in the context of the Kennicutt-Schmidt law \citep{Schmidt1959,Kennicutt1998a}, the central SFEs are comparable to the levels seen in starburst systems, and in conformity with the unresolved observations of AGNs. We discuss the implications of our findings in terms of the effectiveness of AGN feedback and the mass growth of the BHs and host galaxy bulges. 

Section~\ref{sec:obs} summarizes the data and observations. Section~\ref{sec:met} presents the methods to derive the SFR estimates, molecular gas masses, and the spatial distribution of these quantities within the host galaxies. We study in Section~\ref{sec:res} the SFR distributions and examine the SFE relative to that of the normal, star-forming systems. Section~\ref{sec:dis} discusses the implications of our study, and, finally, we summarize in Section~\ref{sec:con}. 

\vskip 1.5cm
\section{Sample and Observations}
\label{sec:obs}

We benefit from archival ALMA and MUSE observations that map at $\sim 0\farcs4$--$1\farcs4$ scales a sub-sample of six $z \lesssim 0.06$ AGN host galaxies extracted from the broader sample of 87 $z < 0.5$ quasars belonging to the PG survey \citep{Boroson1992}. Selected by their optical/ultraviolet colors, the PG quasars are a representative sample of luminous, broad-line (type~1) AGNs unbiased with respect to dust or gas content. This quasar sample is one of the most studied to date, with an available rich repository of multi-wavelength data for the AGN and host galaxy, ranging from X-ray \citep{Reeves2000,Bianchi2009} through optical  \citep{Boroson1992,Ho2009}, mid-IR \citep{Shi2014,Xie2021,Xie2022}, far-IR \citep{Petric2015,Shangguan2018,Zhuang2018}, mm \citep{Shangguan2020,Shangguan2020b}, and radio \citep{Kellermann1989,Kellermann1994} wavelengths, allowing detailed SED modeling and accurate estimation of the global SFR and gas content of the host galaxies \citep{Shangguan2018}. Hubble Space Telescope (HST) $\sim 0\farcs1$ resolution optical and near-IR imaging is also available for a substantual fraction of the sample \citep{Kim2008,Zhang2016,Kim2019,Zhao2021}.

\begin{figure*}
\centering
\includegraphics[width=1.23\columnwidth]{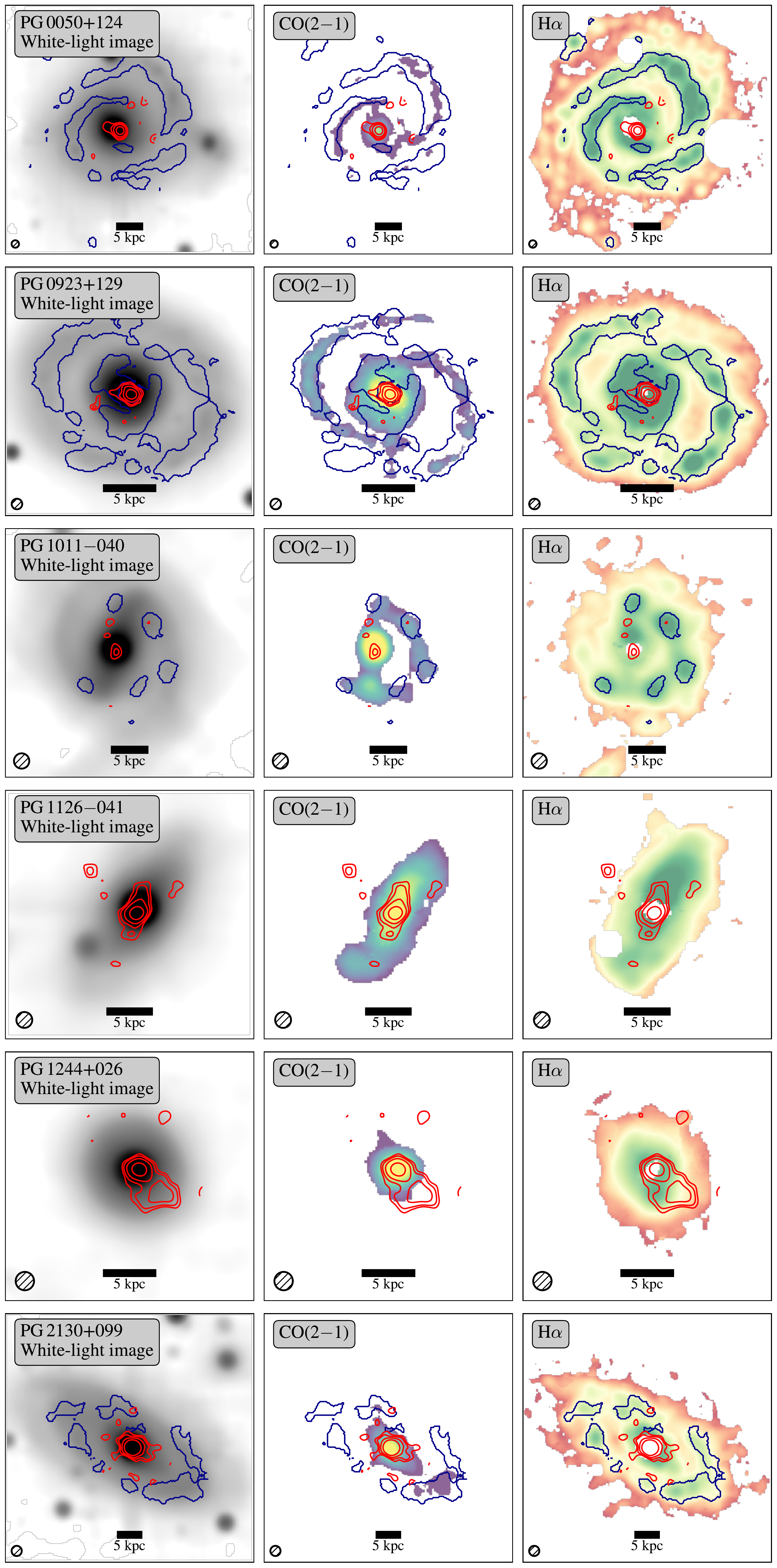}
\caption{\label{fig:Allmaps} MUSE white-light image (\textit{left}), CO(2--1) emission-line intensity map (\textit{middle}), and H$\alpha$ emission-line intensity map (\textit{right}) for the six PG quasar host galaxies. The red contours show the restframe 230\,GHz continuum data at 2, 3, 5, and $10\times$ the observation noise level. The blue contours enclose the ionized gas regions classified as star-forming, if any, following the [N\,{\sc ii}]/H$\alpha$--[O\,{\sc iii}]/H$\beta$ diagnostic diagram \citep{Baldwin1981}. In the bottom-left corner of each panel, we show the effective observation spatial resolution, which corresponds to the ALMA synthesized beam convolved with the MUSE PSF (Section~\ref{sec:SP_matching}).}
\end{figure*}

\subsection{ALMA Observations}
\label{ALMA_obssetup}

The ALMA observations (program 2018.1.00006.S; PI:~F.~Bauer) were presented formally in \citet{Molina2021}. Briefly, those correspond to $\sim 0\farcs4$--$1\farcs4$ resolution Band~6 observations designed to detect the $^{12}$CO($J=2-1$) transition [$\nu_{\rm rest} = 230.538$\,GHz; hereafter CO(2--1)] and the underlying continuum. The data were reduced and calibrated up to the $uv$-products using the standard calibration pipeline within the Common Astronomy Software Application (\texttt{CASA}; \citealt{McMullin2007}) version 5.6.1-8. These ALMA data were concatenated with previous Atacama Compact Array (ACA) observations \citep{Shangguan2020} to maximize the recovery of signal from extended emission. The imaging was performed using \texttt{tclean} with Briggs weighting (\textsc{robust} = 0.5) and the \textsc{auto-multithresh} multi-masking option configured to minimize noise and sidelobe contamination. The observations have a channel resolution of $\sim 11$\,km\,s$^{-1}$.

We further implement two additional steps to regularize the ALMA data with respect to the MUSE data. First, in all the observations we circularize the beam shape by applying a minor convolution within \texttt{CASA}. Secondly, we regrid the ALMA data cubes to match exactly the pixel grid of the MUSE data by using the \texttt{CASA} task \texttt{imregrid}. In the particular case of PG\,0050+124, we repeat the data cleaning process, but this time we impose an output resolution of $1''$ within \texttt{tclean} by using the `uvtaper' option to maximize the recovery of diffuse flux, while closely matching the observation resolution to that of the MUSE data. 

The CO(2--1) intensity (or moment~0) maps were derived by the pixel-wise fitting of the emission lines, employing a multiple Gaussian component modeling procedure, following \citet{Molina2021}. Briefly, a simple Gaussian function is fitted to the spectra and used to determine if an emission line is detected with high enough signal-to-noise (S/N\,$\geq 5$). For some CO emission lines presenting asymmetric or even more complex shapes, we repeated the fit by including up to two additional Gaussian sub-components to the emission-line model. Once a best fit is obtained, the emission-line modeling is repeated several times, but in each case the fit initial guesses are set equal to the model parameter values obtained from the previous fit of a neighboring pixel. We keep the CO(2--1) line model that delivers the lowest Bayesian Information Criterion (BIC; \citealt{Schwarz1978}). This step allows to minimize the algorithm sensitivity to the imposed initial guesses. In each pixel, the best-fit parameter uncertainties are derived using a Monte Carlo resampling method considering 300 iterations (see \citealt{Molina2021} for more details). 

The CO(2--1) luminosity per pixel is calculated following \citep{SV2005}

\begin{equation}
\label{eq:LCO}
L^\prime_{\rm CO(2-1)} = 3.25 \times 10^7\,\frac{S_{\rm CO(2-1)} \Delta v\, D^2_{L}}{\nu_{\rm obs}^{2}\,(1+z)^{3}}\,\,\, {\rm [K\,km\,s^{-1}\,pc^2]},
\end{equation}

\noindent where $S_{\rm CO(2-1)} \Delta v$ is in units of Jy\,km\,s$^{-1}$, $\nu_{\rm obs}$ is the observed frequency of the line in GHz, $D_{L}$ is the luminosity distance in Mpc, and $z$ is the redshift. We estimate CO(1--0) luminosities by adopting the median luminosity ratio $L^\prime_{\rm CO(2-1)} / L^\prime_{\rm CO(1-0)} = 0.62$ value found by \citet{Shangguan2020} for PG quasars at $z < 0.3$. We estimate molecular gas masses adopting a CO-to-H$_2$ conversion factor $\alpha_{\rm CO} = 3.1$\,$M_\odot$\,(K\,km\,s$^{-1}$\,pc$^2$)$^{-1}$ with 0.3\,dex uncertainty \citep{Sandstrom2013}, a value consistent with dust-based gas masses independently derived for the PG quasars \citep{Shangguan2020}. The CO(2--1) intensity maps are presented in Figure~\ref{fig:Allmaps}. 

We note that the absolute astrometric uncertainty of our ALMA data is $\lesssim 0\farcs05-0\farcs07$ ($\lesssim 0.25-0.35 \times$\,pixel scale), given the high S/N detection of our targets,\footnote{https://almascience.nao.ac.jp/documents-and-tools/cycle8/alma-technical-handbook} which implies that the ALMA data do not require astrometric correction.

\subsection{MUSE Observations}
\label{sec:MUSE_obssetup}

The MUSE observations of six PG quasar host galaxies were presented previously in \citet{Molina2022}. They correspond to MUSE wide-field-mode seeing-limited and ground-layer adaptive optics (AO)-aided observations that deliver $\sim 0\farcs8$--$1\farcs4$ resolution maps of the host galaxies over a field-of-view (FoV) of $\sim 1' \times 1'$ with a pixel sampling size of $0\farcs2 \times 0\farcs2$. The spectra cover the wavelength range $\sim$4700--9350\,\r{A} with 1.25\,\r{A}\,pixel$^{-1}$ at a mean resolution of $R \approx 3000$, or full width at half maximum (FWHM) $\sim 2.65\,$\r{A}. The MUSE observations were carried out under the European Southern Observatory (ESO) programs 094.B$-$0345(A), 095.B$-$0015(A), 0103.B$-$0496(B), and 0104.B$-$0151(A). The data cubes were further processed by applying the Zurich Atmosphere Purge (ZAP) sky subtraction tool \citep{Soto2016} to optimize the removal of residual sky-subtraction features. We corrected the spectra for Galactic reddening assuming the \citet{Cardelli1989} extinction law and the extinction values of \citet{Green2019}.

In each MUSE data cube, the AGN emission is subtracted from that of the host galaxy using the AGN emission deblending methodology presented in  \citet{Molina2022}. This procedure is based on the fact that, for each quasar observation, the nuclear spectrum is effectively observed as a blurred point-like source that follows the PSF across the data cube. Upon modeling the nuclear spectrum following the approach for characterizing broad-lined type~1 AGN spectra (e.g., \citealt{Greene2005b}), the resulting derived nuclear spectral template is used as an additional sub-component when pixel-wise fitting the spectra encoded in the data cubes (see \citealt{Molina2022} for details). In this work, we rederive the nebular emission-line maps for each quasar host galaxy after cross-matching the spatial resolution of the ALMA and MUSE observations (Section~\ref{sec:SP_matching}). We show the H$\alpha$ intensity maps in Figure~\ref{fig:Allmaps}. The astrometry of the MUSE data is corrected by aligning the spectrally collapsed images (``white-light images'') with respect to images from the Sloan Digital Sky Survey (SDSS; \citealt{Abolfathi2018}), if available, or else from the Panoramic Survey Telescope and Rapid Response System (PanSTARRS; \citealt{Chambers2016}), using the MUSE \textsc{Python} Data Analysis Framework (\texttt{MPDAF}) routine \texttt{estimate\_coordinates\_offset} \citep{Bacon2017}. Only in the case of PG\,1244+026 is the \texttt{MPDAF} routine not able to find any solution when aligning both images, and for this particular case we only correct the MUSE data astrometry by matching the quasar coordinates.

\begin{table*}
	\centering
	\def\arraystretch{1.2}
	\setlength\tabcolsep{3pt}
    	\caption{\label{tab:cont_prop} Restframe 230\, GHz Continuum Properties}
    	\vspace{0.2mm}
	\begin{tabular}{ccccccccccccc}
		\hline
		\hline
		Object & Beam size & $S_{230}^{\rm map}$ & Best model & $S_{230}^{\rm ps}$ & $S_{230}^{\rm ext}$ & $R_{\rm e, 230}$ & $b/a$ & PA & $f_{\rm AGN}$ & $f_{\rm ff}$ & $f_{\rm syn}$ & $\alpha_{\rm R}$\\
		& ($''$) & (mJy) & & (mJy) & (mJy) & ($''$) & & ($^\circ$) & (\%) & (\%) & (\%) & \\
		(1) & (2) & (3) & (4) &(5) & (6) & (7) & (8)& (9) & (10) & (11) & (12) & (13) \\
		\hline
	PG\,0050+124 & 1.0 & $1.06\pm0.05$ & Gau.  & \nodata & $0.87\pm0.06$ & $0.1\pm0.1$ & 0.8 & 124 & $<1$ & 24 & 4 & $-0.91$ \\
         PG\,0923+129 & 1.2 & $0.87\pm0.04$ & Ps.\,+\,Exp. & $0.35\pm0.07$ & $1.25\pm0.28$ & $2.5\pm0.7$ & 1.0 & \nodata & $1$ & 23 & 40 & $-0.61$ \\
         PG\,1011$-$040 & 1.3 & $0.08\pm0.01$ & Ps. & $0.1\pm0.04$ & \nodata & \nodata & \nodata & \nodata & $>8$ & $>34$ & $>23$ & $-0.70$ \\
         PG\,1126$-$041 & 1.1 & $0.44\pm0.02$ & Ps.\,+\,Gau. & $0.2\pm0.02$ & $0.56\pm0.17$ & $2.7\pm0.6$ & 0.64 & 35 & $2$ & 28 & 10 & $-0.65$ \\
         PG\,1244+026 & 1.3 & $0.71\pm0.02$ & Ps.\,+\,Gau. & $0.3\pm0.04$ & $0.33\pm0.14$ & $1.7\pm0.6$ & 0.24 & 43 & $2$ & 6 & 4 & $-0.84$ \\
         PG\,2130+099 & 1.4 & $0.80\pm0.04$ & Ps.\,+\,Gau. & $0.4\pm0.06$ & $0.35\pm0.10$ & $1.0\pm0.3$ & 0.85 & 72 & $1$ & 10 & 11 & $-0.81$ \\
		\hline
	\end{tabular}
	\justify
	{\justify \textsc{Note}--- (1) Source name. (2) Circularized synthesized beam size (FWHM) of the ALMA observation. (3) Total continuum flux density at restframe 230\,GHz. (4) Best-fit model chosen by BIC: ``Ps.'' = point-like source; ``Gau.'' = Gaussian; ``Exp.'' = exponential. (5) Continuum flux density of the point-like source component, if any. (6) Continuum flux density of extended component, if any. (7) Half-light radius associated with the extended continuum component. (8) Continuum axial ratio measured by \texttt{uvmodelfit}. (9) Position angle of the major axis of the continuum distribution measured with respect to the north in anti-clockwise direction. (10) Contribution from AGN dust heating to the restframe 230\,GHz continuum flux density, estimated estimated by adopting the models detailed in \citet{Shangguan2018}. (11) Estimated thermal free-free emission contribution at restframe 230\,GHz. (12) Contribution from synchrotron emission to the continuum measurements. (13) Spectral index used to model the synchrotron emission SED component. We note that we have not considered the ALMA flux calibration uncertainty ($\lesssim 10$\,\%; \citealt{Fomalont2014,Bonato2018}) when reporting the continuum flux density values.}
\end{table*}

\section{Methods}
\label{sec:met}

Our main goal is to study the distribution of SFR and SFE within the host galaxies. We need to carefully match the resolution of the ALMA and MUSE observations, as well as to control properly for any AGN emission contamination when deriving the SFRs from the optical and far-IR data. Once the unresolved AGN emission is deblended from the host galaxy, we further control for AGN photoionized gas emission before using the H$\alpha$ fluxes to compute the SFRs. When doing this, we explore different ``mixing-sequence'' corrections (e.g., \citealt{Wild2010,Davies2014a,Davies2014b,Davies2016}). The far-IR continuum data are converted to SFRs using the panchromatic SED models for the PG quasars of \cite{Shangguan2018} and the global SFR$_{\rm IR}$ estimates of \cite{Xie2021}. The latter quantity is also adopted to assess the degree to which the mm continuum flux density is contaminated by thermal free-free emission, while we use complementary radio data to evaluate the effect of synchrotron emission from an AGN jet. The total SFRs traced by the optical emission are quantified and compared with the global values of SFR$_{\rm IR}$, providing an estimate of the nuclear star formation missed in the optical light data due to inaccurate AGN-deblending and/or central dust attenuation. The following explains the analysis steps in detail.

\subsection{Matching the Spatial Resolution}
\label{sec:SP_matching}

While the ALMA and MUSE observations have nearly the same spatial resolution in terms of FWHM ($\sim 0\farcs8-1\farcs4$), the ALMA synthesized beam and MUSE PSF have different radial profiles. The ALMA synthesized beam is Gaussian-shaped, while the MUSE PSF is well-described by a \cite{Moffat1969} function. To match the spatial resolution of both data sets, we simply convolve the ALMA and MUSE observations by the PSF and synthesized beam, respectively. For each observation, the ALMA beam is constructed by using the data cube header information, while the MUSE PSF is derived as part of the AGN emission deblending technique of \citet{Molina2022} to the MUSE data cubes before applying any convolution.

\begin{figure*}
\centering
\includegraphics[width=1.8\columnwidth]{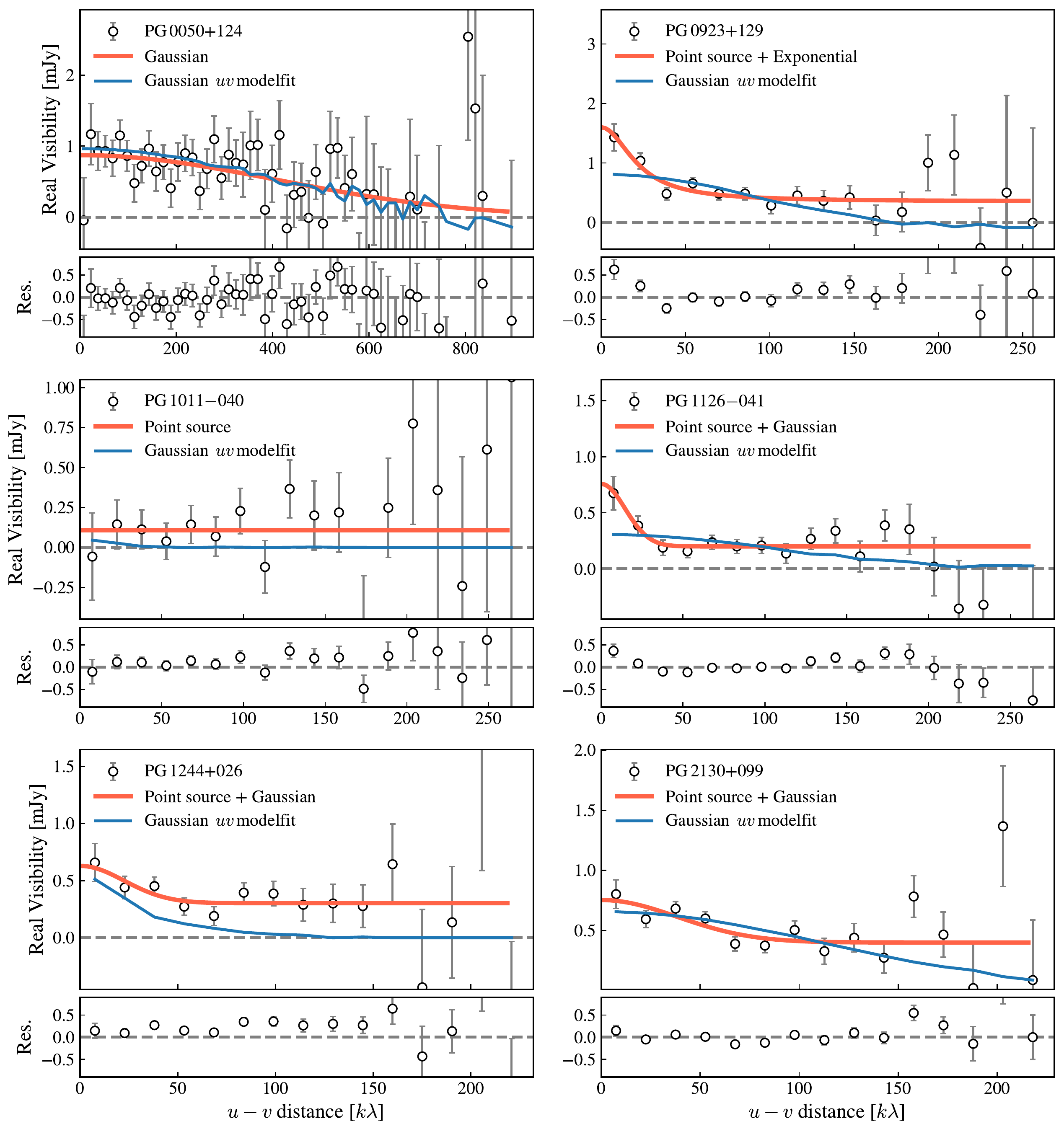}
\caption{\label{fig:uvcontfits} Real part of the visibilities as a function of $u-v$ distance for the PG quasar hosts. The red line shows the best-fit model to the observed profile accordingly to BIC. The blue line represents the best-fit Gaussian model derived by \texttt{uvmodelfit} to the full visibility data, and the lower panel shows the residuals from this model. The continuum is resolved on $\sim$kpc scales in five out of the six host galaxies.}
\end{figure*}

\subsection{Characterizing the Millimeter Continuum}
\label{sec:Cont_model}

The restframe 230\,GHz mm continuum traced by ALMA in Band 6 provides critical information in support of the MUSE data to deliver a complete mapping of the on-going star formation activity. For each host galaxy, we model the visibility data using \texttt{uvmodelfit} in \texttt{CASA}, by adopting a simple two-dimensional Gaussian model. Additionally, we extract the profiles of the collapsed real part of the visibilities in terms of $u-v$ distance binned in intervals of 15\,k$\lambda$ from the calibrated measurement sets. In each $u-v$ distance bin, we compute the average real part of the visibilities, and we estimate the uncertainty from the standard deviation of the data. We use \texttt{lmfit} \citep{Newville2014} to fit three basic model profiles: (1) point source, (2) Gaussian profile, and (3) exponential profile.  We explore two additional options combining the profile models with a point-source sub-component, a Gaussian profile plus a point source, and an exponential profile plus a point source. We evaluate the fit that best describes the visibility data by employing BIC analysis.  Results are given in Table~\ref{tab:cont_prop}. Figure~\ref{fig:uvcontfits} shows the visibility profiles and their best fits, along with the profile of the models derived by \texttt{uvmodelfit} to the entire visibility data. We note that the continuum size estimates are reliable given the high S/N ($\sim13-40$) of the continuum peak for the systems with detected extended emission \citep{Simpson2015}.

We aim to use the continuum observations to map the obscured SFR activity, so for the sake of consistency we compute the total restframe 230\,GHz continuum flux density from the continuum images.  For each source,  we sum the continuum emission from individual pixels with S/N\,$> 2$ within a radial aperture of $6 ''$ centered on the continuum peak. The aperture is set wide enough to enclose all the source emission detected in the ALMA continuum maps, but avoiding the noisy data typically seen at the edges of the data cubes. With the exception of the systematic uncertainty of the ALMA flux calibration ($\lesssim 10$\,\%; \citealt{Fomalont2014, Bonato2018}), the $1\,\sigma$ error of the total fluxes is dominated by the adopted aperture size. We vary the aperture value between $5''$ and $9''$ in steps of 0$\farcs$5 and re-estimate the total fluxes. The $1\,\sigma$ error is taken from the standard deviation of those values.  We note that those values broadly agree with the total fluxes estimated from the visibility data modeling (Table~\ref{tab:cont_prop}).

\begin{figure*}
\centering
\includegraphics[width=2.0\columnwidth]{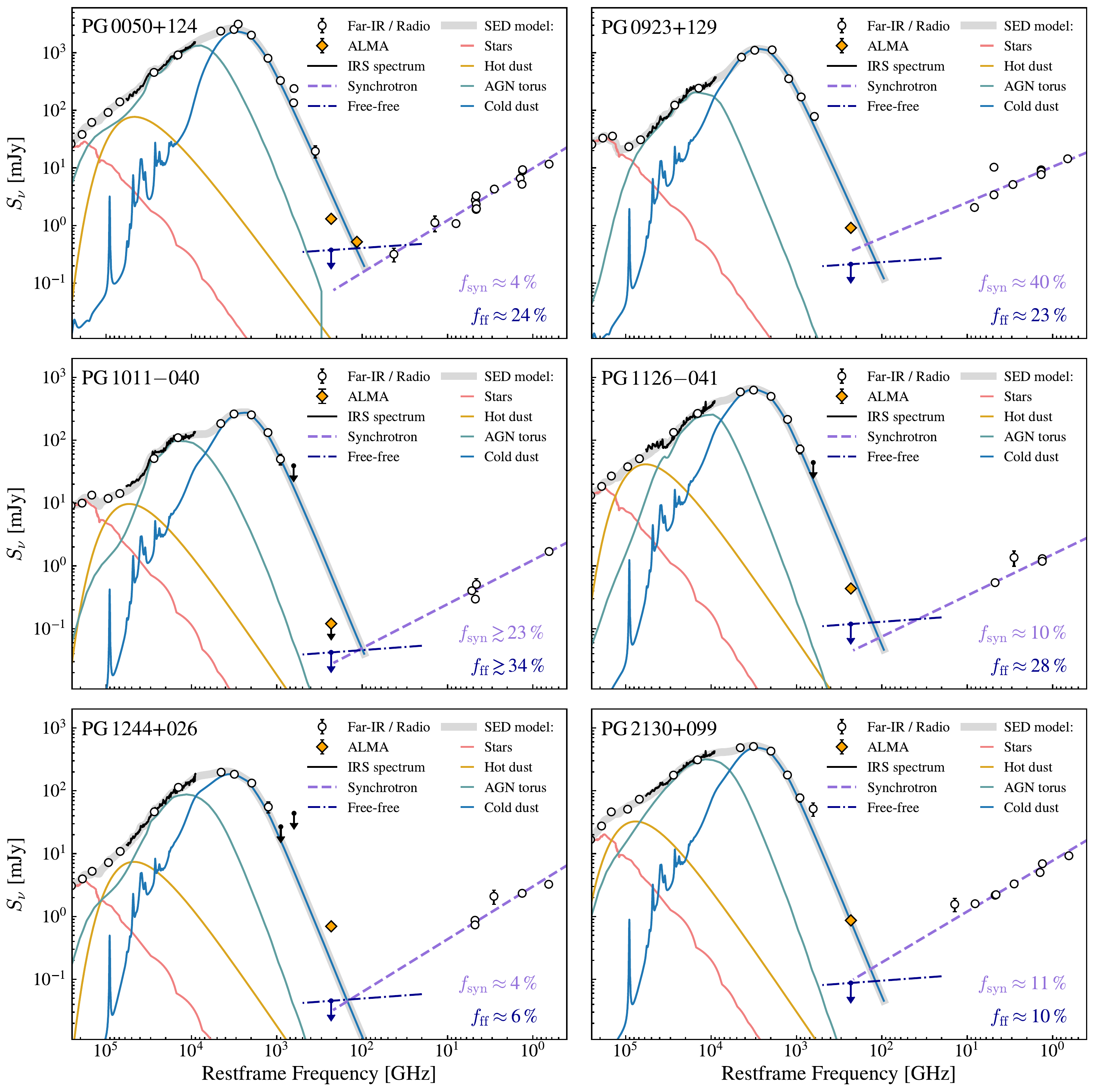}\\
\caption{\label{fig:submmSED} Far-IR SED of the six PG quasar host galaxies. The open circles show the IR data from 2MASS, WISE, and Herschel collated in \citet{Shangguan2018}, along with the GHz radio data taken from the literature (Table~\ref{tab:radio_data}). The orange diamond indicates the ALMA mm continuum at restframe $\sim 230$\,GHz. In the case of PG\,0050+124, we also report the ALMA mm continuum at restframe $\sim 115$\,GHz \citep{Tan2019}. Upper limits are indicated by downward arrows. The black solid line represents the available IRS spectrum in the $\sim 5- 40\,\mu$m wavelength range. We show the best-fit SED model in grey band, while the colored solid lines represent the model sub-components (see details in \citealt{Shangguan2018}): starlight (red), hot dust (yellow), AGN torus (green), and cold dust (blue). The dot-dashed line shows the upper limit of the thermal free-free emission estimated from the total SFR$_{\rm IR}$ of the host galaxy. The dashed line indicates synchrotron emission, extrapolated assuming a power-law shape ($S_\nu \propto \nu^{\alpha_{\rm R}}$) with index estimated from the 1.4 and 5\,GHz flux densities. For PG\,1011$-$040, we assume $\alpha_{\rm R} = -0.7$ because of the lack of 1.4\,GHz data. The contribution of synchrotron ($f_{\rm syn}$) and free-free ($f_{\rm ff}$) emission to the measured restframe $230$\,GHz continuum flux density are listed in the bottom-right corner of each panel. For PG\,1011$-$040, we provide a lower limit to $f_{\rm syn}$ and $f_{\rm ff}$ by adopting the upper limit of the mm continuum flux density.}
\end{figure*}

\subsection{Cleansing the Millimeter Continuum Flux Density}
\label{sec:Cont_decont}

Before using the ALMA mm continuum data to estimate SFRs, we need to remove additional emission components that might arise from the AGN torus, synchrotron emission from a jet, or free-free emission from thermal plasma associated with the central engine. The expected AGN torus emission is estimated following the best-fit SED model provided by \citet{Shangguan2018}. They constructed a panchromatic IR SED using photometry from $\sim 1$ to $500\,\mu$m, complemented by mid-IR ($5-38\,\mu$m) spectra from the Spitzer Infrared Spectrometer (IRS). The IR SED was modeled using physically motivated emission components for the starlight, AGN dust torus, and host galaxy dust. To help constrain the synchrotron component, we use radio data from the literature taken at lower frequencies ($\sim 0.7-15$\,GHz; Appendix~\ref{sec:AppA}). We adopt a synchrotron power-law spectrum with fixed spectral index $\alpha_{\rm R}$ ($S_\nu \propto \nu^{\alpha_{\rm R}}$), which is computed from the observed flux densities at 1.4 and 5\,GHz, except for PG\,1011$-$041, for which we set $\alpha_{\rm R} = -0.7$ \citep{Smolcic2017}, because it only has radio data at a single frequency. In cases where many flux density measurements are available, we use the average value as reference.  To estimate the thermal free-free emission, we follow the SFR-based conversion given by \citet{Murphy2011} and the values of the total SFR$_{\rm IR}$ (Table~\ref{tab:sample}). This conversion assumes the initial mass function (IMF) of \citet{Kroupa2001}. We show in Figure~\ref{fig:submmSED} the global SEDs and the best-fit SED models presented in \citet{Shangguan2018}. Note that while we did not use the full radio SED to estimate $\alpha_{\rm R}$, the extrapolation of the synchrotron component provides reasonable description of all the radio measurements. 

Recently, \citet{Kawamuro2022} found that the compact ($\lesssim 100\,$pc scale) mm emission detected in local BASS AGN hosts is correlated with the AGN X-ray flux. They suggest that the mm emission may correspond to self-absorbed synchrotron radiation around the AGN X-ray corona. \citet{Kawamuro2022} provided a best-fit trend between this compact mm emission with the AGN X-ray fluxes (their Table 1), which we use to predict the corresponding flux densities for the PG quasar hosts by adopting the available soft X-ray (2--10\,keV) data \citep{Bianchi2009,Ricci2017b}. We find that the predicted mm emission is in reasonable agreement (i.e., under the 0.45\,dex scatter reported by \citealt{Kawamuro2022}) with our observationally supported flux density estimates mainly associated with synchrotron and thermal free-free emissions, which is encouraging given the adoption of global SED to estimate the flux density unassociated with dusty star formation activity in AGNs. This suggests that we are successfully subtracting this contamination source from our restframe 230\,GHz continuum emission measurements. 

For each source, we compute contamination levels attributed to the AGN torus ($f_{\rm AGN}$), synchrotron ($f_{\rm syn}$), and thermal free-free ($f_{\rm ff}$) emission by simply taking the ratio between the predicted flux density at restframe 230\,GHz with the total continuum level measured from the ALMA observation.

\begin{figure*}
\centering
\includegraphics[width=1.4\columnwidth]{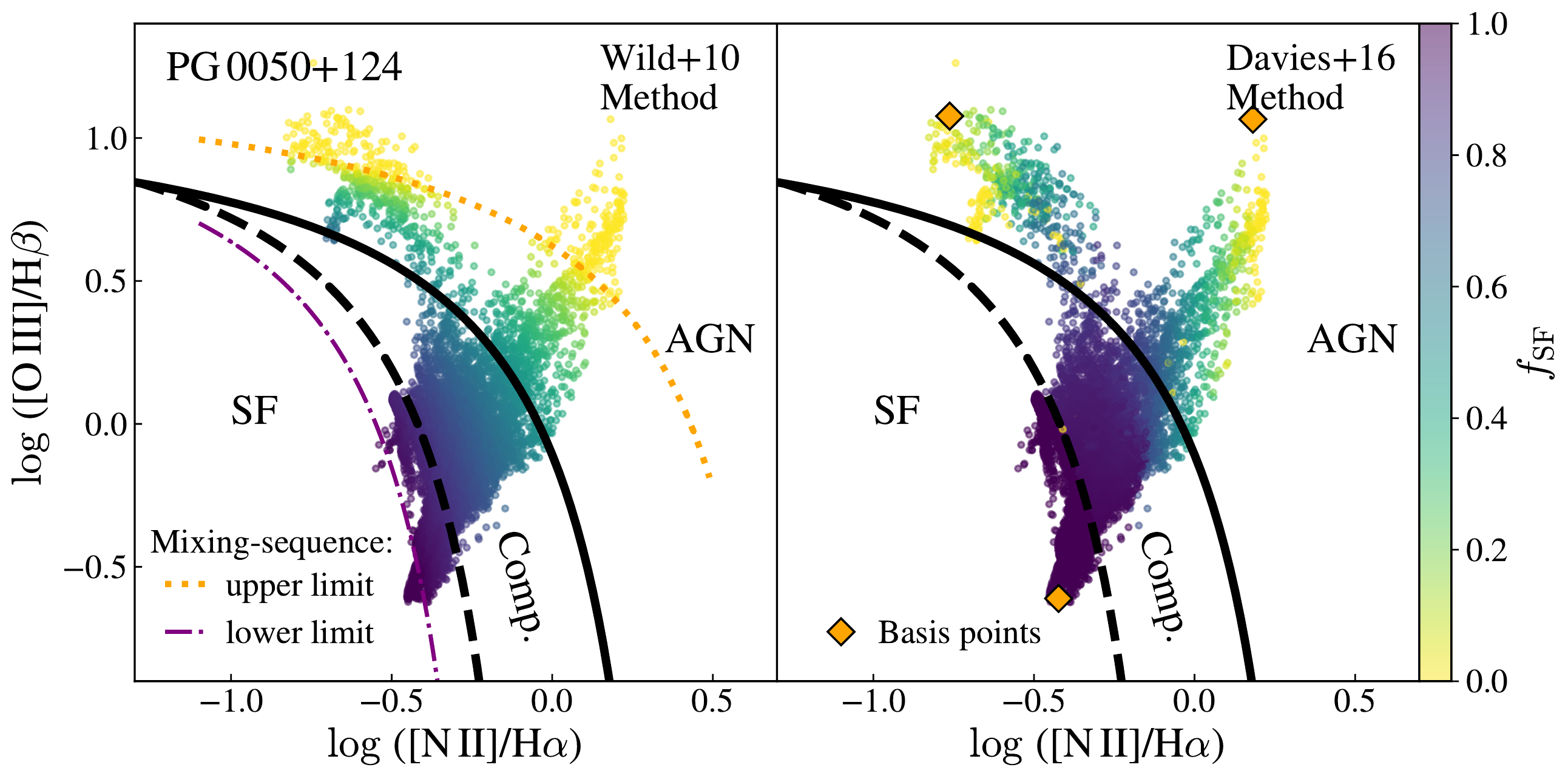}\\
\includegraphics[width=1.4\columnwidth]{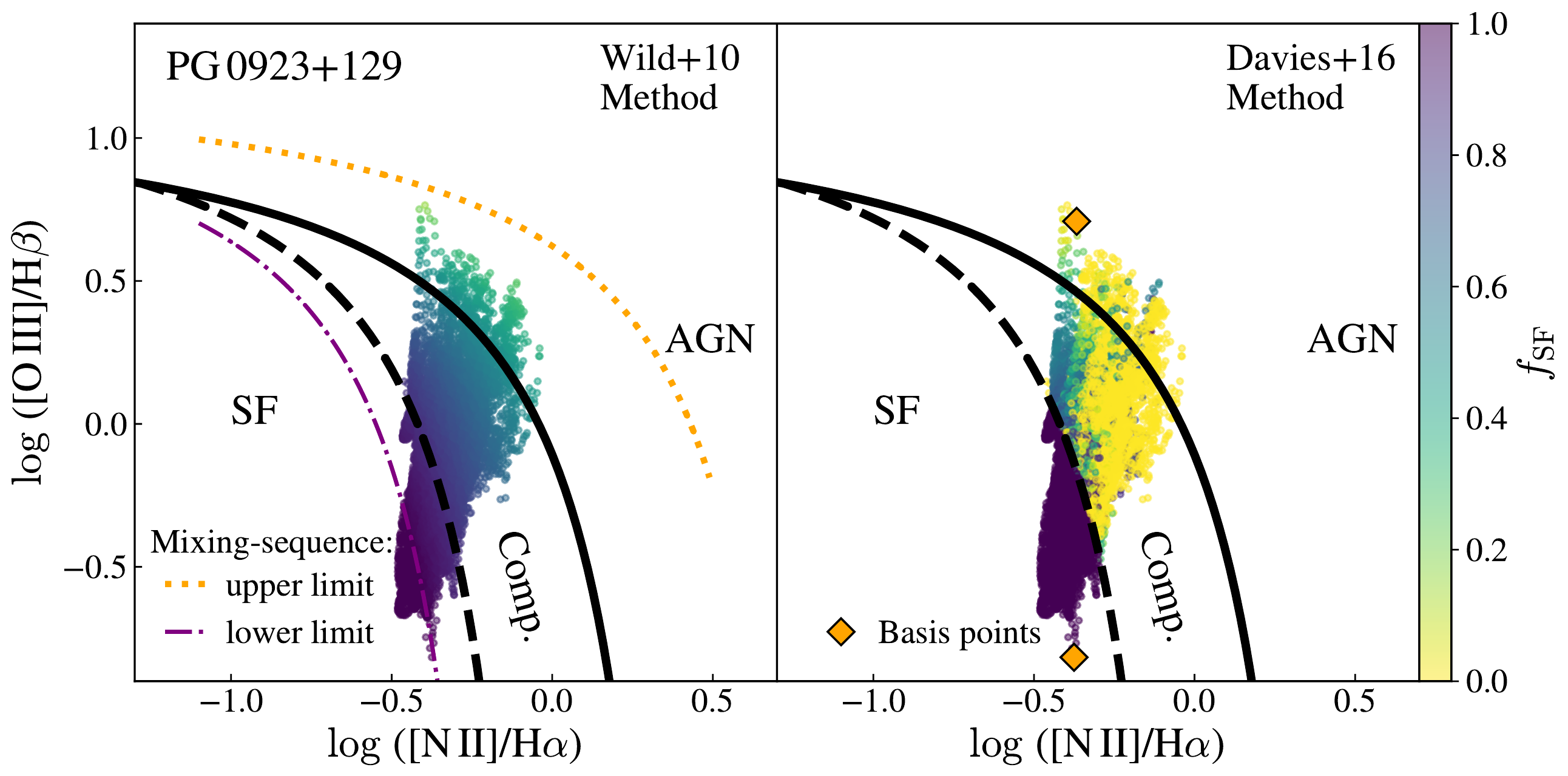}\\
\includegraphics[width=1.4\columnwidth]{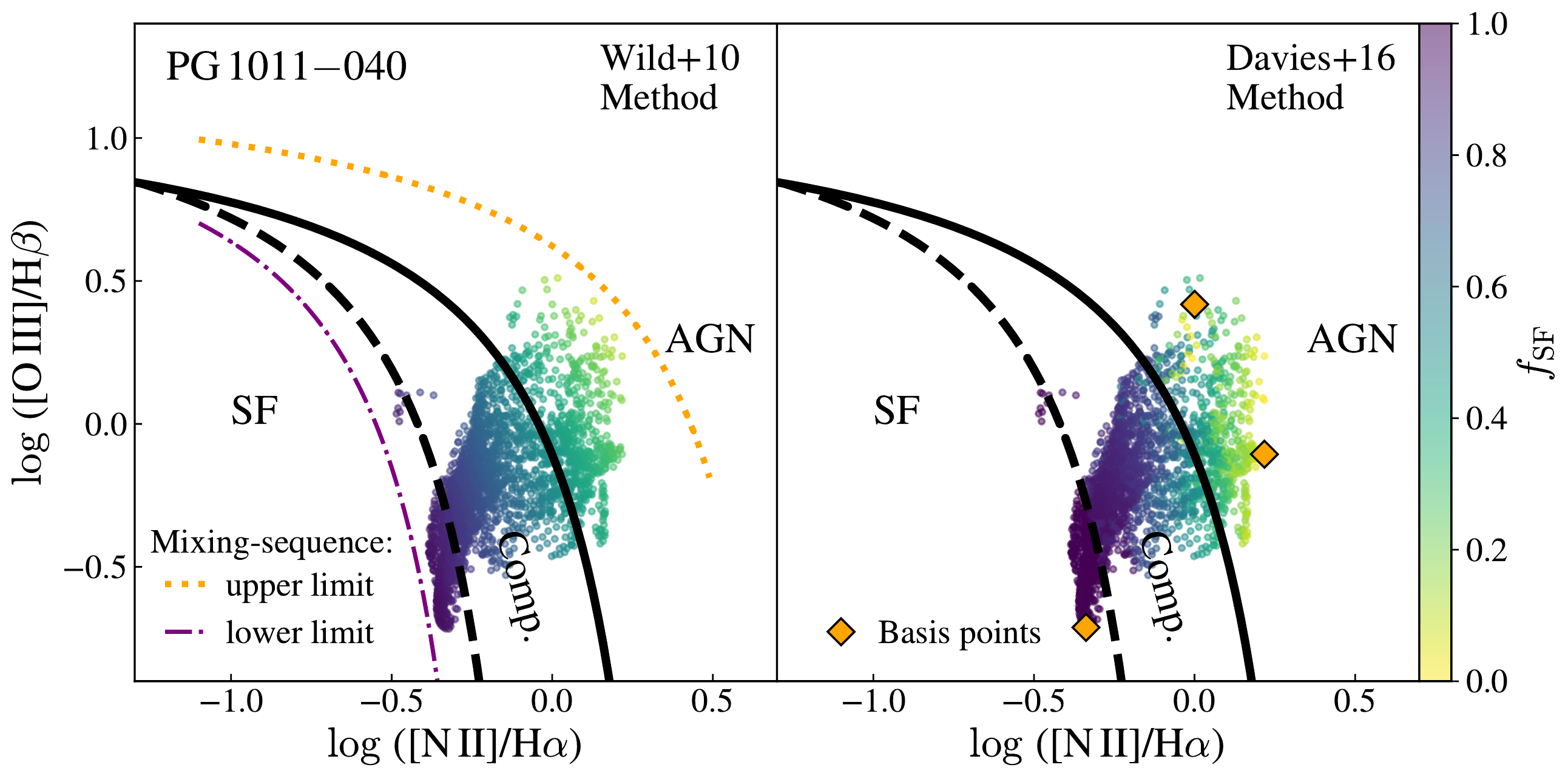}\\
\caption{\label{fig:mixing_comp} BPT diagram for the PG quasar host galaxies, highlighting the different H$\alpha$ flux correction factors derived from each mixing-sequence method. The left column shows the mixing-sequence method of \citet{Wild2010}. The dotted curve marks the upper limit of the mixing-sequence zone, while the dot-dashed curve indicates its lower extent; above the former curve, the line emission is assumed to be produced solely by AGN radiation, while the opposite is adopted for the data lying below the latter curve. The right column shows our modified version of the mixing-sequence method developed by \citet{Davies2016}, if applicable. The orange diamonds highlight the basis points used when deriving the values of $f_{\rm SF}$. All panels are color-coded by $f_{\rm SF}$. In each panel we also show the maximum starburst limit (solid curve; \citealt{Kewley2001}) and the demarcation for pure star-forming systems (dashed curve; \citealt{Kauffmann2003}).}
\end{figure*}

\begin{figure*}
\centering
\includegraphics[width=0.825\columnwidth]{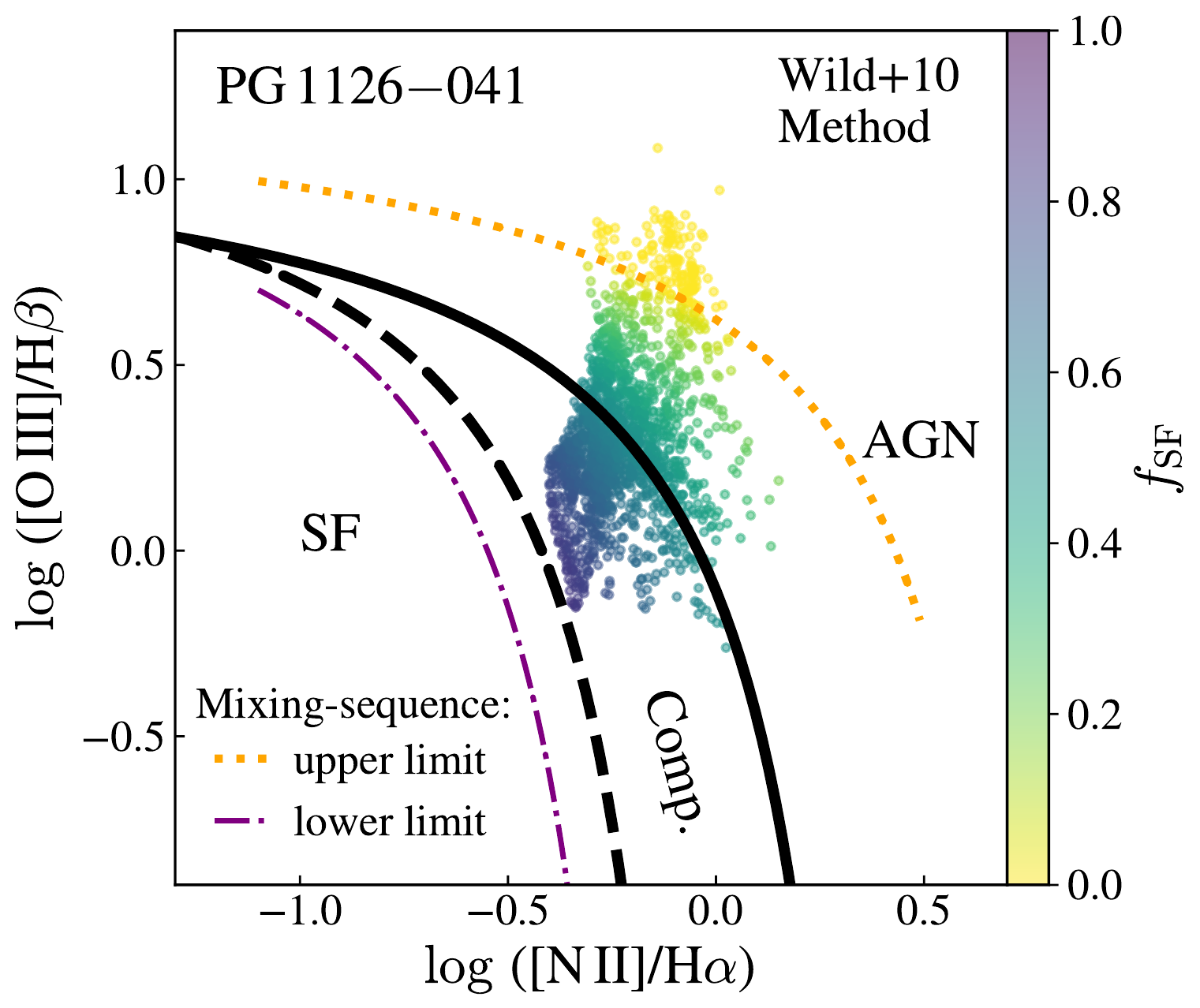}\\
\includegraphics[width=0.825\columnwidth]{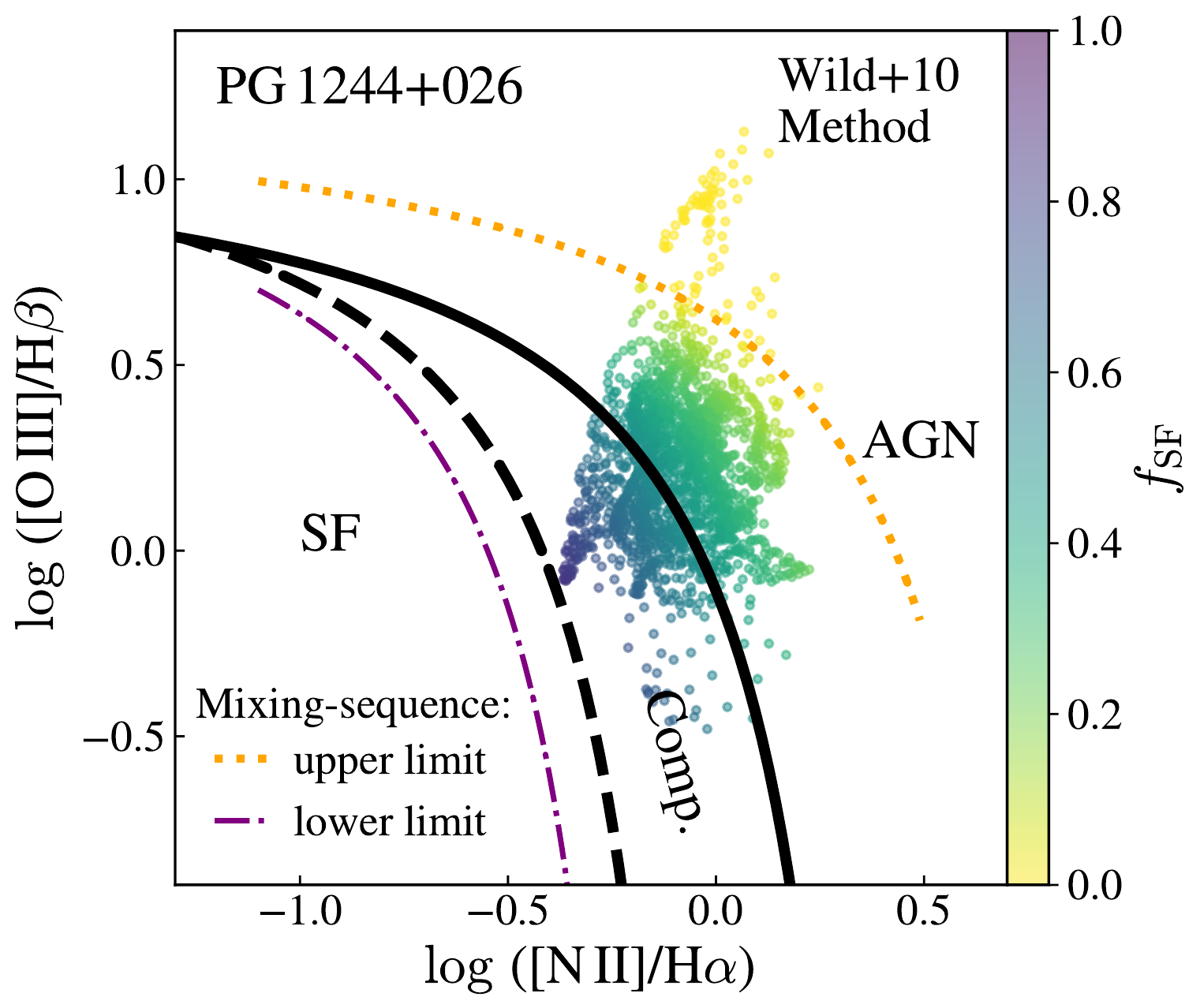}\\
\includegraphics[width=1.4\columnwidth]{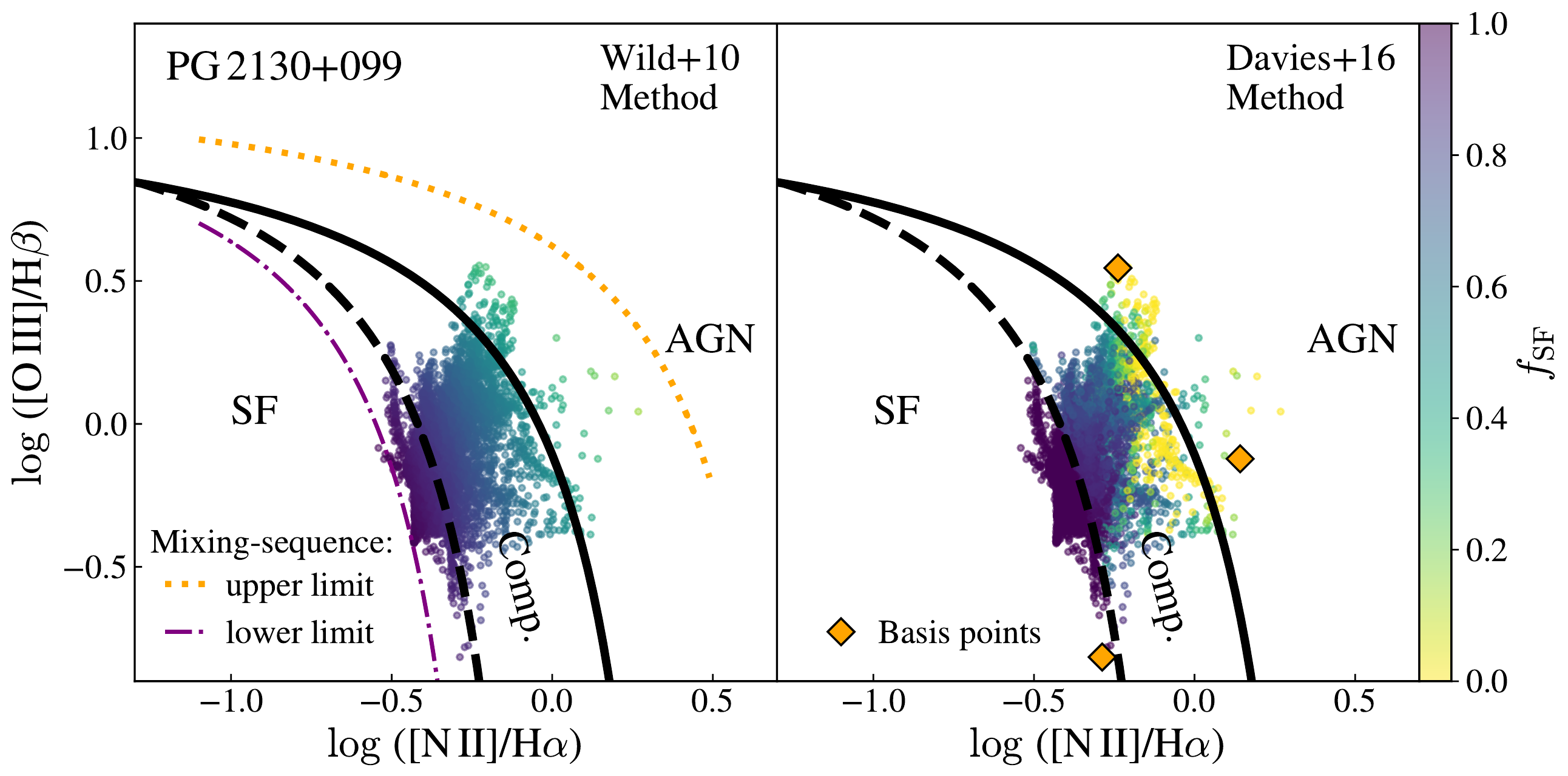}\\
\centering{\textbf{Figure~\ref{fig:mixing_comp}}. Continued.}
\end{figure*}

\subsection{Mixing-sequence Correction}
\label{sec:MS_method}

The standard calibrations used to estimate SFRs cannot be applied directly in active galaxies. The AGN-deblending procedure only removes the point-like AGN emission from the observations, while the AGN photoionization effect on the H$\alpha$ line-emitting gas still needs to be corrected for. To address this issue, mixing-sequence corrections have been developed. The main assumption of the method is that the emission-line fluxes measured across the line-of-sight correspond to a combination of the underlying emission associated with star formation and that produced by the AGN photoionization (e.g., \citealt{Davies2014a,Davies2014b,Davies2016,Wylezalek2018,Husemann2019b}; but see \citealt{Agostino2021} for a different interpretation). As there are various mixing-sequence correction methods in the literature, we adopt two different procedures to isolate the H$\alpha$ emission produced by star formation activity (e.g., \citealt{Smirnova-Pinchukova2022}). We stress that our main objective is not to determine the best procedure, but to assess the uncertainty associated with adopting any method of this kind. We note that the emission source classification of each pixel is done using the [N\,{\sc ii}]/H$\alpha$--[O\,{\sc iii}]/H$\beta$ diagnostic diagram (\citealt{Baldwin1981}; hereinafter BPT) with the traditional maximum starburst \citep{Kewley2001} and pure star-forming \citep{Kauffmann2003} demarcation curves. We only consider the pixels in which the emission lines are detected with S/N\,$\gtrsim 3$. The emission-line fluxes were corrected for dust attenuation using the Balmer decrement, assuming an intrinsic ratio of H$\alpha$/H$\beta = 2.86$ for star-forming pixels and 3.1 for others \citep{Osterbrock2006}, and adopting the dust attenuation curve of \citet{Calzetti2000}.

The first method that we apply corresponds to that outlined in \citet{Wild2010}. Characterizing galaxy-integrated SDSS spectra in the [N\,{\sc ii}]/H$\alpha$--[O\,{\sc iii}]/H$\beta$ BPT diagram, they define two new demarcation curves following the parameterization of the traditional pure star-forming curve of \citet{Kauffmann2003}, but shifted in [O\,{\sc iii}]/H$\beta$ ratio. The data lying below the lower demarcation curve are assumed to trace regions in which 100\,\% of the emission is produced by star formation, while the data above the other demarcation curve represent zones fully ionized by the AGN. The data that lie between both curves correspond to composite emission (see their Figure~6). For a single data point, $f_{\rm SF}$ is given by the ratio of the orthogonal distances to both curves (their Equation~2). 

The second method that we adopt is the mixing-sequence correction method developed by \citet{Davies2016}. Briefly, for each host galaxy, a mixing sequence is derived in the [N\,{\sc ii}]/H$\alpha$--[O\,{\sc iii}]/H$\beta$ diagram by fitting a linear function to the data. The extremal points (in terms of [O\,{\sc iii}]/H$\beta$) closer to the mixing sequence define the AGN and star-forming ``basis points'' that encompass the respective [O\,{\sc iii}], H$\alpha$, [N\,{\sc ii}], [S\,{\sc ii}] emission-line flux sets ({\bf F}). The regions classified as star-forming in the BPT diagram are considered to be free from AGN contamination, while the other data are assumed to reflect a linear flux combination of both basis points, 

\begin{equation}
\label{eq:mix_seq}
{\bf F}_{\rm obs} = f_{\rm SF} \times {\bf F}_{\rm SF} + (1 - f_{\rm SF}) \times {\bf F}_{\rm AGN}, 
\end{equation}

\noindent so that $f_{\rm SF}$ can be computed by pixel-wise fitting the emission-line flux sets \citep{Davies2014a,Davies2014b,Davies2016}. One major drawback of this methodology is that the mixing-sequence is undefined if there is no star-forming classified pixel in the data (e.g., PG\,1126$-$041), meaning that no correction can be applied. Another issue is that galaxies are often represented by [N\,{\sc ii}]/H$\alpha$--[O\,{\sc iii}]/H$\beta$ diagrams that are not well-characterized by two simple basis points. This is the case, for example, when the gas-phase metallicity varies across the galaxy, or if a low-ionization nuclear emission-line region (LINER; e.g., \citealt{Ho2008b}) is present. To alleviate this problem, we apply a modified version of this procedure in which we relax the constraint of using only two basis points by adding more basis points to the pool (e.g., \citealt{Husemann2019a}). For a particular pixel, $f_{\rm SF}$ is then given by the combination of the star-forming and AGN basis point that better represents the emission-line flux dataset (i.e., the fit that provides the lowest $\xi^2$ value) among all the possible basis point combinations. In our case, 2--3 basis points per galaxy are usually required to estimate the pixel-wise $f_{\rm SF}$ values. 

Figure~\ref{fig:mixing_comp} shows the [N\,{\sc ii}]/H$\alpha$--[O\,{\sc iii}]/H$\beta$ diagram color-coded by $f_{\rm SF}$ for the PG quasar host galaxies. The color code in each panel highlights the different $f_{\rm SF}$ values provided by each routine. We compute a pixel-wise data scatter of 0.17\,dex between both methods. However, in terms of the total H$\alpha$ flux attributed to star formation activity, the scatter reduces to 0.04\,dex among our targets. The reason for such agreement is that both methods weigh similarly the star formation-classified pixels, namely the pixels that predominantly contribute to the H$\alpha$ emission associated with star formation (see Appendix~\ref{sec:AppC} for more details). We add this uncertainty to our error budget, and hereafter we proceed by only adopting the mixing-sequence correction of \citet{Wild2010} to estimate the SFR from the H$\alpha$ emission. 

\begin{table}
	\centering
	\def\arraystretch{1.2}
	\setlength\tabcolsep{2pt}
    	\caption{\label{tab:SFR_vals} Spatially Integrated SFRs}
    	\vspace{0.2mm}
	\begin{tabular}{cccccc}
		\hline
		\hline
		Object & SFR$^{\rm D16}_{\rm H\alpha}$ & SFR$^{\rm W10}_{\rm H\alpha}$ & SFR$_{230}$ & SFR$_{\rm cen}$\\
		& ($M_\odot$\,yr$^{-1}$) & ($M_\odot$\,yr$^{-1}$) & ($M_\odot$\,yr$^{-1}$) & ($M_\odot$\,yr$^{-1}$)\\
		(1) & (2) & (3) & (4) & (5) \\
		\hline
	PG\,0050+124 & 4.81 & 4.25 & $5.43$ & 22.05 \\
	PG\,0923+129 & 1.31 & 1.31 & $0.51$ & 2.09 \\
        PG\,1011$-$040 & 0.47 & 0.48 & \nodata & 2.42 \\
        PG\,1126$-$041 & \nodata & 0.36 & $2.75$ & 8.34 \\
        PG\,1244+026 & \nodata & 0.04 & $7.06$ & 2.06 \\
        PG\,2130+099 & 0.61 & 0.56 & $5.45$ & 6.54 \\
		\hline
	\end{tabular}
	\justify
	{\justify \textsc{Note}--- (1) Source name. (2) and (3) SFR estimated using, respectively, the AGN-decontaminated H$\alpha$ line fluxes after applying our modified version of the mixing-sequence correction of \citet{Davies2016} and the procedure outlined in \citet{Wild2010}. For the H$\alpha$-based SFRs, the $1\,\sigma$ uncertainty inherent to the mixing-sequence correction method is 0.04\,dex (Appendix~\ref{sec:AppC}). (4) SFR estimated from the restframe 230\,GHz dust continuum emission, which has $1\,\sigma$ uncertainty of 0.3\,dex. (5) ``Central'' SFR estimated following SFR$_{\rm IR} - $\,SFR$_{\rm H\alpha}$, which has a typical uncertainty of $\sim 0.2-0.3\,$dex \citep{Xie2021}. }
\end{table}

\subsection{Star Formation Rate Estimates}
\label{sec:SFR_recipes}

Once the far-IR and H$\alpha$ fluxes have been AGN-decontaminated, we safely infer the SFRs using standard calibrations and provide SFR maps. We estimate the unobscured SFR from H$\alpha$ using \citet{Kennicutt1998b}'s calibration, renormalized to a \citet{Kroupa2001} IMF \citep{Calzetti2013}:

\begin{equation}
\label{eq:SFRHa}
{\rm SFR_{H \alpha}} \,(M_\odot \,{\rm yr^{-1}})= 5.5 \times 10^{-42} L_{\rm H \alpha} \,({\rm erg \, s^{-1}}). 
\end{equation}

\noindent To calculate the obscured SFR from the 230\,GHz continuum (SFR$_{\rm 230}$), we estimate the fraction of flux density detected in the maps compared to the total amount given by the global SED model (Figure~\ref{fig:submmSED}). Then, this fraction is multiplied by the total IR SFR previously estimated by \citet{Shangguan2018} using consistent SFR calibration and IMF assumption (Table~\ref{tab:sample}).\footnote{We note that \citet{Xie2021} corroborated that the far-IR emission correlates with the SFRs derived independently from the mid-IR neon fine-structure lines \citep{Zhuang2019}, demonstrating that the far-IR emission effectively traces young stars in quasar host galaxies.}  Here, we implicitly assume that the central (nuclear) far-IR SED is similar to the global far-IR SED, following the reports for X-ray selected AGN hosts at $z < 0.05$ \citep{Mushotzky2014}. \citet{Xie2021} estimated a $1\,\sigma$ uncertainty of $\sim 0.2-0.3$\,dex for SFR$_{\rm IR}$, and we add this uncertainty to the SFR$_{230}$ error budget (Table~\ref{tab:SFR_vals}).

\subsection{Radial Profiles}
\label{sec:rad_prof}

We characterize the SFR and molecular gas distribution by constructing azimuthally averaged profiles. The inclination angle of the host galaxy is measured from the minor-to-major axis ratio estimated using \texttt{photutils} \citep{Bradley2020} from the MUSE white-light image. The thickness of the tilted ring is set equal to half of the observation spatial resolution. For each host galaxy, the pixels with undetected emission are masked, and the data belonging to each ring are simply averaged.  We explore two S/N pixel masking thresholds, S/N$\,= 3$ and 1. For the latter case, we also adopt a data replacement strategy, where the masked values are replaced by 0 and the data noise level. Thus, we make a total of 3 radial profiles per quantity. For each ring, the uncertainty is computed by the data standard deviation and normalized to the number of independent data points, following Appendix~E of \citet{Leroy2008}. We further mask the rings with completeness (calculated over the non-masked pixels) below 20\,\%.

\section{Results}
\label{sec:res}

The main goal of this work is to study the star formation activity and its distribution in quasar host galaxies. We present a qualitative view in Figures~\ref{fig:Allmaps}, where the blue contours highlight the regions classified as ``star-forming'' following the [N\,{\sc ii}]/H$\alpha$--[O\,{\sc iii}]/H$\beta$ diagram analysis applied to the MUSE observations. We find that the star formation activity traced by the H$\alpha$ emission is ubiquitous and occurs within spiral arms (e.g., PG\,0050+124, PG\,2130+099), H\,{\sc ii} region complexes (e.g., PG\,1011$-$040), inner rings (e.g, PG\,0923+129), and ``composite'' regions (e.g., PG\,1126$-$041, PG\,1244+026; Figure~\ref{fig:mixing_comp}). Any line detection toward the nuclear zones of the host are limited by the accuracy of the quasar deblending procedure. 

The star-forming regions are largely co-spatial with the [CO(2--1)] molecular gas component, in qualitative agreement with that found for inactive systems \citep{Leroy2008}. On the other hand, the 230\,GHz restframe continuum emission---dominated by cold thermal dust emission---is resolved by ALMA (Figure~\ref{fig:uvcontfits}). The AGN-decontaminated dust emission mainly traces the ongoing star formation in the central region of the host galaxy, although the continuum flux detection at large galactocentric radii is currently severely limited by sensitivity (see Appendix~\ref{sec:AppD} for detailed comments for each source). The dust continuum and H$\alpha$ maps furnish complementary views of the SFR distribution, and together they offer a more complete view of the ongoing star formation activity in these systems. We now proceed to analyze the spatially resolved data in detail.

\subsection{Spatially Integrated Star Formation Rates}
\label{sec:SI_SFR}

The H$\alpha$ emission provides a new, independent estimate of the SFR compared to previous values derived from modeling of the global IR SED (Table~\ref{tab:sample}). We find spatially integrated SFR$_{\rm H\alpha}$ values that account for $\sim 2\,\%-39\,$\% of the global SFR$_{\rm IR}$ estimates. This result is independent of the mixing-sequence method employed to remove the AGN contamination (Appendix~\ref{sec:AppC}), although we note that the SFR$_{\rm H\alpha}$, which is sensitive to flux over-subtraction in our quasar deblending procedure, should be considered lower limits to the global SFRs. Nevertheless, adopting ${\rm SFR}_{\rm cen} = {\rm SFR}_{\rm IR} - {\rm SFR}_{\rm H\alpha}$ provides a lower limit to the ongoing star formation activity in the central region of the host galaxy (Table~\ref{tab:SFR_vals}).

From the restframe 230\,GHz continuum maps, we compute spatially integrated SFR values that encompass $\sim 15\,\%-80$\,\% of the global SFR$_{\rm IR}$, where the larger differences stem from systems with higher flux density corrections for synchrotron and thermal free-free emissions ($\gtrsim 30$\,\% of total emission at 230\,GHz when accounted together). However, the uncertainty associated with applying both corrections is small compared to that given by scaling the global far-IR SED model to a single continuum data point when computing SFR$_{230}$ ($\sim 0.2-0.3$\,dex). The synchrotron flux density contamination is typically $\lesssim 10$\,\%, with the exception of PG\,0923+129 where we estimate $f_{\rm syn} \approx 40$\,\%. Considerable uncertainty surrounds our estimates of $f_{\rm syn}$, which derive from heterogeneous and non-simultaneous radio data from the literature (Figure~\ref{fig:submmSED}; Table~\ref{tab:radio_data}). If we adopt $\alpha_{\rm R} = -0.7$ \citep{Smolcic2017}, SFR$_{230}$ would vary by 3\,\%--40\,\%. On the other hand, the contribution from free-free emission,  $f_{\rm ff} = 6\,\%-28\,\%$, could be reduced by a factor of 2 if we account for the dust absorption of Lyman continuum \citep{Yun2002}, an effect that would increase SFR$_{230}$ by 7\,\%--30\,\%. 

PG\,1244+026 is a striking case where we measure SFR$_{230} > \,$SFR$_{\rm IR}$. Such a glitch can be understood partly by considering that we adopt a fixed dust emissivity index for the far-IR SED within the \citet{DraineLi2007} models (roughly a fixed far-IR spectral index $\beta \approx 2.08$) adopted to fit the global photometric data \citep{Shangguan2018}. This SED modeling systematic may also affect the accuracy of SFR$_{\rm IR}$ for the other host galaxies. For example, if we assume a $\beta$ variation from 2.08 to 1.5 \citep{Bianchi2013}, then the predicted flux density at 230\,GHz restframe would be increased by a factor of 9; however, in such a case, both SFR$_{230}$ and SFR$_{\rm cen}$ would highly disagree. After excluding PG\,1244+026, the scatter between both SFR$_{230}$ and SFR$_{\rm cen}$ is 0.22\,dex. Fitting the far-IR SED using different models or templates (e.g., \citealt{Jones2013,Ciesla2014}) would provide further insights, as well as considering the effect of efficient dust heating by young stars \citep{Nersesian2019}, but this is beyond the scope of this work. It is noteworthy that PG\,1244+026 exhibits an asymmetric component in its dust continuum (Figure~\ref{fig:Allmaps}) that is co-spatial with a low-S/N emission feature at 0.685\,MHz \citep{Silpa2020}, although it is not detected in CO emission (Figure~\ref{fig:Allmaps}). We speculate that the far-IR excess may be related to a dusty, diffuse, cold-gas outflow component that is completely outshined in the Herschel bands.

For the systems with estimates of both SFR$_{230}$ and SFR$_{\rm H\alpha}$, we find that the sum of those values account for 16,\,\%--84\,\% of the global values of SFR$_{\rm IR}$, with an average of 45\,\% (after excluding PG\,1244+026, due to its anomalous flux continuum measurement). In light of the significant systematic uncertainties involved in each quantity, we consider them to be in fair agreement.

\begin{figure*}
\centering
\includegraphics[width=2.0\columnwidth]{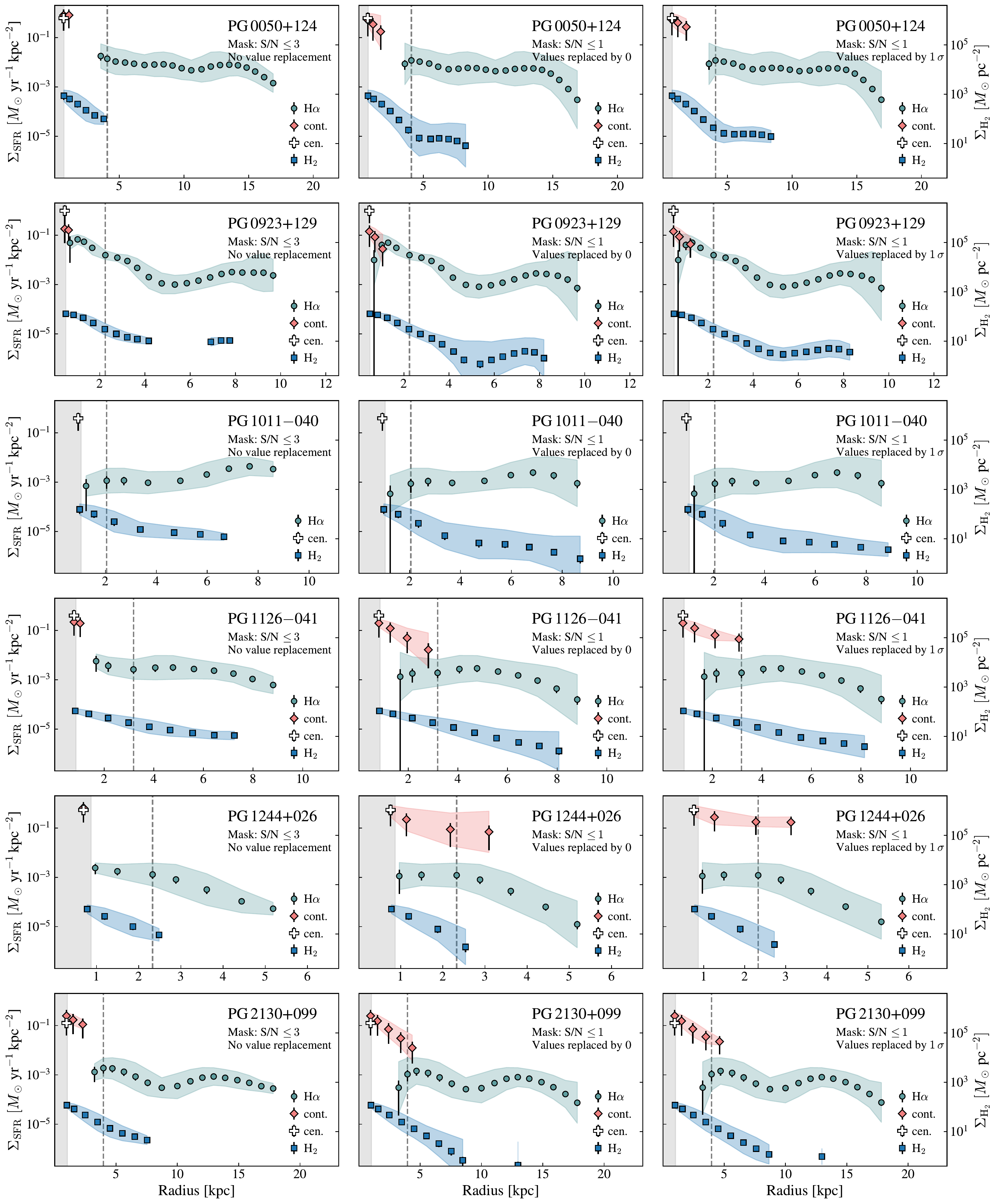}
\caption{\label{fig:radial_profiles} Star formation rate surface density radial profiles for the six PG quasar host galaxies (rows).  We derive three different radial profiles per quantity for each host galaxy considering different pixel data S/N threshold levels and replacement values (columns).  The radial profiles are derived from both AGN-decontaminated H$\alpha$ and restframe 230\,GHz continuum. We also show the average host galaxy central $\Sigma_{\rm SFR}$ computed as SFR$_{\rm cen} / A$, with $A$ corresponding the host galaxy area where we miss the optical-light emission due to AGN deblending flux oversubtraction. We present the $\Sigma_{\rm H_2}$ radial profiles with the corresponding scaling in the right $y$-axis. The grey-shaded zone indicates the spatial resolution of the observations. The other color-shaded zones represent the respective pixel-wise standard deviation. In each panel, the dashed line delimits the central region where the majority ($>$\,50\%) of the MUSE pixels tend to be masked due to AGN emission over-subtraction, and indicates uncertain $\Sigma_{\rm SFR}$ measurements at smaller galactocentric radius.}
\end{figure*} 

\subsection{Star Formation Activity and Molecular Gas Distribution}
\label{sec:SFR_profiles}

To analyze and compare the star formation activity and molecular gas distributions, we construct surface density SFR ($\Sigma_{\rm SFR}$) and $M_{\rm H_2}$ ($\Sigma_{\rm H_2}$) radial profiles. We compute separate profiles for the SFR maps based on dust continuum and H$\alpha$ (Figure~\ref{fig:radial_profiles}), as they trace the obscured and unobscured star formation activity over different time scales \citep{Kennicutt2012}.  

We find a large variety of H$\alpha$-based $\Sigma_{\rm SFR}$ profiles toward the host galaxy centers: increasing trends (PG\,0923+129, PG\,1126$-$041, and PG\,1244+026), flat trends (PG\,0050+124 and PG\,2130+099), and even a decreasing trend (PG\,1011$-$040). However, we note that those profiles are affected by different systematics at large and small radius. At large radii, the profiles are affected by detection bias, in that the measured values mainly reflect the emission coming from the spirals arms (e.g., PG\,0050+124, PG\,2130+099), outer H\,{\sc ii} region complexes (e.g., PG\,1011$-$040), or simply the density flux detection threshold imposed and/or the data replacement value.  Closer to the quasar location,  the AGN outshines the optical light of the host galaxy. The quasar deblending procedure fails to separate both spectral sub-components, which results in the oversubtraction of the host galaxy emission.  The H$\alpha$-based $\Sigma_{\rm SFR}$ radial profiles are only computed from a handful of points, each of which is uncertain because of the need for AGN decontamination.  It should be also considered that the central zone of the PG quasar hosts presents large molecular gas surface densities ($\gtrsim 100\,M_\odot$\,yr$^{-1}$; \citealt{Molina2021}),  with possible severe dust attenuation ($A_V \gtrsim 3$ mag; \citealt{Barrera-Ballesteros2020}) limiting the recovery of the H$\beta$ emission-line flux, hence the intrinsic $\Sigma_{\rm SFR}$ profile.\footnote{The H$\beta$ emission is difficult to measure in zones where $A_V \gtrsim 3$ mag and may significantly underestimate $A_V$ for regions with $A_V \gtrsim 2$ \citep{Liu-G2013}.} The dashed lines in Figure~\ref{fig:radial_profiles} serve as rough reference to indicate the region where the H$\alpha$-based $\Sigma_{\rm SFR}$ is significantly underestimated. 

As a key difference with the H$\alpha$ observations, the dust continuum-based $\Sigma_{\rm SFR}$ profiles trace the innermost obscured, central star formation activity. Figure~\ref{fig:radial_profiles} shows the dust-based profiles, with the exception of PG\,1011$-$040, whose central dust continuum is very weak and unresolved. We find that the dust continuum-based measurements, with $\Sigma_{\rm SFR} \approx 0.15-0.80$\,$M_\odot$\,yr$^{-1}$\,kpc$^{-2}$, are systematically higher than those derived from H$\alpha$.  Note that most of the objects present a sub-component of unresolved continuum emission (Figure~\ref{fig:uvcontfits}).  Coupled to the fact that we have adopted conservative factors to correct for synchrotron and thermal free-free emission, the reported values of $\Sigma_{\rm SFR}$ are likely lower limits.  This result is independent of the method applied to derive the radial profiles,  but we caution that constraining the continuum-based $\Sigma_{\rm SFR}$ radial profile down to the SFR surface density levels measured in H$\alpha$ emission is uncertain and limited by the sensitivity of our observations.  The radial profiles constructed by masking all the pixels with S/N$\,\leq 3$ are more affected by detection bias. Only when including the low-S/N data are we able to map fainter levels of $\Sigma_{\rm SFR}$ from the continuum emission.  Those profiles tend to agree better in their overlap with the H$\alpha$-based $\Sigma_{\rm SFR}$ profiles (with the exception of PG\,1244+026,  the host galaxy with an asymmetric component in its dust continuum map and anomalous mm total flux). However,  the maps in the outer region are patchy, implying that the treatment of the pixels with undetected emission severely affects the azimuthal average estimate at a given radius.  For instance, in Figure~\ref{fig:radial_profiles} we also show the radial profile for each PG quasar host extracted by including fainter emission (S/N$\geq1$) and adopting two data replacement values---zero and the data $1\,\sigma$ level, the two cases that should embody the true radial profile shape.  For the former case,  the continuum-based $\Sigma_{\rm SFR}$ profiles decline more steeply with radius, while in the latter case the radial profiles tend to flatten at large radius,  converging to the data $1\,\sigma$ level.  This can be more easily appreciated in the cases of PG\,1126$-$041 and PG\,2130+099.  We note that PG\,0923+129 is the only system for which a straightforward extrapolation of the H$\alpha$-based $\Sigma_{\rm SFR}$ radial profile would match the values traced by the dust emission. Interestingly, this system is also the one observed at the highest projected spatial resolution ($\sim 1.0\,$kpc), given its low redshift ($z = 0.03$), and thus its H$\alpha$-based $\Sigma_{\rm SFR}$ profile is less affected by the systematics of the AGN-deblending procedure. For the other systems, the combination of the MUSE and ALMA data suggests a steep rise of $\Sigma_{\rm SFR}$ toward the center.  This result is also supported by the rough consistency between the dust continuum-based $\Sigma_{\rm SFR}$ and that obtained by dividing SFR$_{\rm cen}$ by the deprojected area of the central region where the H$\alpha$ and H$\beta$ line emissions are oversubtracted by the AGN deblending procedure and/or may be affected by severe dust attenuation.    

We show, for reference, the $\Sigma_{\rm H_2}$ radial profiles. Unlike the H$\alpha$-based $\Sigma_{\rm SFR}$ radial profiles, $\Sigma_{\rm H_2}$ increases toward small radius in all the sources. Indeed, with the exception of PG\,1126$-$041, whose CO(2--1) half-light radius is $3.8$\,kpc \citep{Molina2021}, for the rest of the sample most ($\gtrsim 50$\,\%) of the molecular gas mass located within the central $\sim 2$\,kpc. Recall that the molecular gas data correspond to a concatenation of ALMA 12-m array and ACA data aiming to maximize the CO(2--1) emission line recovery (Section~\ref{ALMA_obssetup}). The concatenated data show good agreement with ACA data alone in terms of measured total CO(2--1) fluxes, indicating that the molecular gas maps trace most of the emission coming from the host galaxies (see \citealt{Molina2021} for a more detailed analysis).

\begin{figure*}
\centering
\includegraphics[width=2.0\columnwidth]{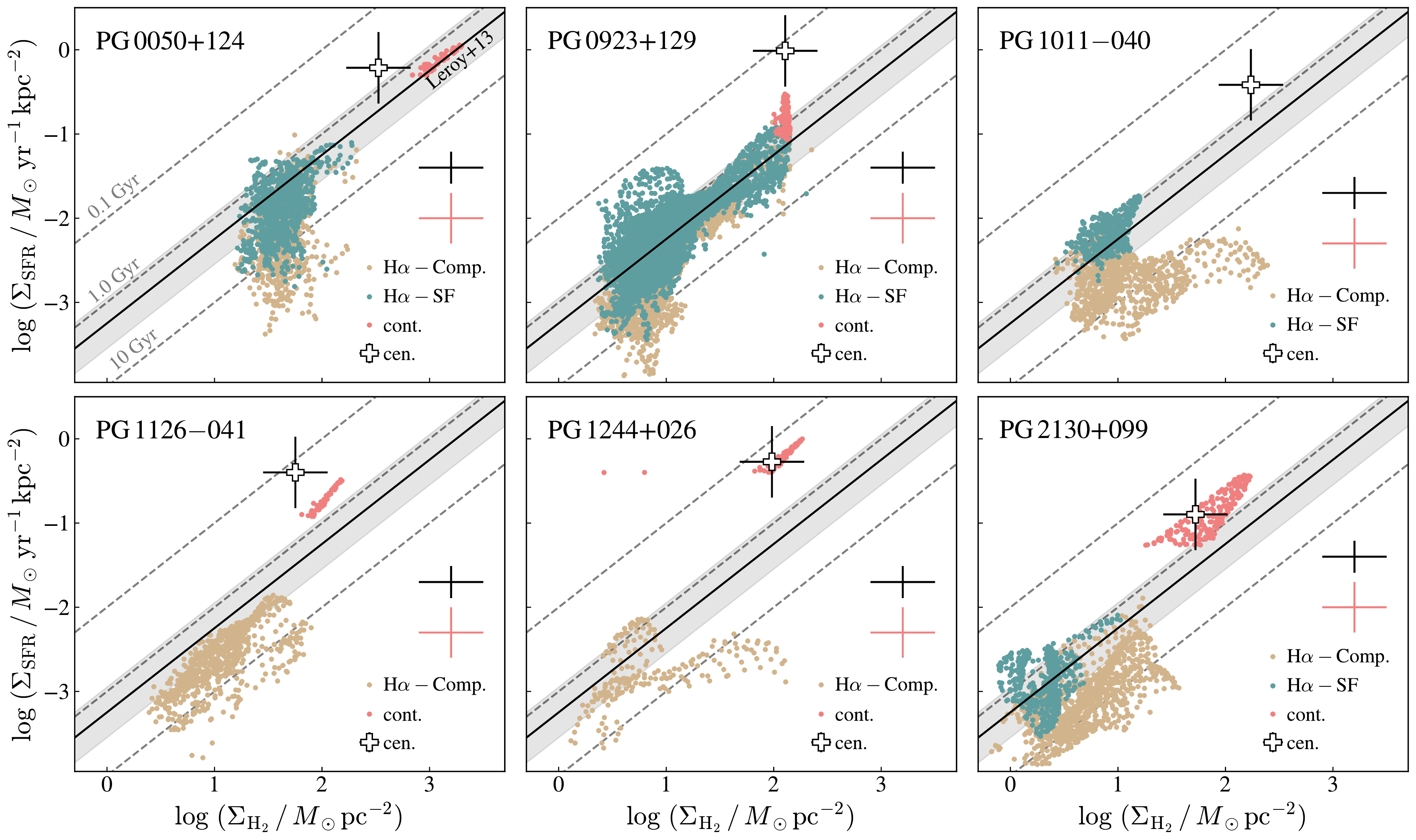}
\caption{\label{fig:KS_plot_indv} The PG quasar host galaxies in the Kennicutt-Schmidt diagram \citep{Kennicutt1998a}. The red points show the $\Sigma_{\rm SFR}$ values estimated from the restframe 230\,GHz continuum data. We show in green dots the MUSE pixels classified as star-forming, while the other pixels are represented by brown dots.  The plus symbol presents the average host galaxy central value computed over the host galaxy area where we miss the optical-light emission due to AGN deblending flux oversubtraction. The solid line shows the best-fit values of $t_{\rm dep}$ derived by \citet{Leroy2013} for local inactive galaxies, with the shaded region corresponding to the $1\,\sigma$ uncertainty. The dashed lines represent values of constant $t_{\rm dep}$, as labelled in the upper-left panel. The regions with obscured and unobscured star formation activity are characterized by $t_{\rm dep}$ values similar to or slightly higher than those measured from the inactive systems, indicating starburst-like activity.  The black and red colored error bars located above the legend correspond to the typical $1\,\sigma$ uncertainty of the H$\alpha$- and continuum-based pixel-wise data. }
\end{figure*}

\subsection{The Kennicutt-Schmidt Relation}
\label{sec:KS_relation}
The star formation activity follows a power-law relation between the gas mass and SFR surface density, $\Sigma_{\rm SFR} \propto \Sigma_{\rm gas}^N$ \citep{Schmidt1959,Kennicutt1998a}. In local spiral galaxies, global measurements find that this empirical Kennicutt-Schmidt relation has a power-law index $N\approx1.4$ \citep{delosReyes2019}, but when spatially resolving the galaxies and only considering the molecular gas component ($\Sigma_{\rm gas} = \Sigma_{\rm H_2}$), largely $N\approx1.0$ (e.g, \citealt{Leroy2008}). The common interpretation is that the molecular gas is converted into stars at a constant star formation efficiency (SFE\,$\equiv \Sigma_{\rm H_2}/\Sigma_{\rm SFR}$) or gas depletion time ($t_{\rm dep} \equiv 1/{\rm SFE} \approx 1.8$\,Gyr; \citealt{Leroy2013}).\footnote{This estimate corresponds to $t_{\rm dep} = 2.2$\,Gyr as reported by \citet{Leroy2013} after adjusting to the CO-to-H$_2$ conversion factor, IMF, and CO(2--1)/CO(1--0) ratio adopted in this work.} More active star-forming galaxies or extreme starbursts systems such as ultra/luminous infrared galaxies (U/LIRGs) are characterized by higher SFEs (shorter $t_{\rm dep}$). When normal star-forming galaxies and starburst systems are analyzed together, the power-law index increases to $\sim 1.5$ \citep{Kennicutt2021}. Using the Kennicutt-Schmidt relation as a benchmark framework, \citet{Shangguan2020b} already showed, on the basis of globally average measurements of $\Sigma_{\rm SFR}$ and $\Sigma_{\rm H_2}$, that PG quasars have SFEs higher than those of normal, star-forming galaxies and instead are more akin to starburst systems. Here we delve deeper into this topic using spatially resolved observations.

Figure~\ref{fig:KS_plot_indv} shows the PG quasar hosts in the context of the Kennicutt-Schmidt relation. We compare the data in terms of gas depletion time, while avoiding the data with $\Sigma_{\rm H_2} \lesssim 12 \,M_\odot$\,pc$^{-2}$, where the gas surface density is mainly determined by atomic hydrogen \citep{Leroy2008}, and focusing on $\Sigma_{\rm SFR} > 0.001\,M_\odot\,$yr$^{-1}$\,kpc$^{-2}$, wherein H$\alpha$ is a reliable tracer of the star formation activity \citep{Kennicutt2012}. The MUSE data are labeled in terms of the pixels classified following the conventional [N\,{\sc ii}]/H$\alpha$--[O\,{\sc iii}]/H$\beta$ diagnostic diagram, according to the degree of mixing-sequence correction (see Section~\ref{sec:MS_method}). Recall that the ALMA and MUSE datasets are matched in terms of spatial resolution and pixel size. We find that the regions classified as star-forming tend to show $t_{\rm dep} \approx 1.8$\,Gyr, while we measure higher $t_{\rm dep}$ values for the remaining zones, a result not particularly sensitive to the mixing-sequence correction method adopted in this work. Nevertheless, we find somewhat significant variations among the host galaxies. In PG\,0050+124 and PG\,0923+129, the H$\alpha$-based $\Sigma_{\rm SFR}$ estimates scatter around the $t_{\rm dep} \approx 1.8$\,Gyr estimate; however, we find slightly higher $t_{\rm dep}$ for others (e.g., PG\,1126$-$041, PG\,2130+099). Careful scrutiny indicates that the shorter values of $t_{\rm dep}$ based on H$\alpha$ are found in H\,{\sc ii} region complexes located mainly at the galaxy outskirts and within spiral arm-like substructures (see Figure~\ref{fig:Allmaps}). The higher H$\alpha$-based $t_{\rm dep}$ estimates tend to be found in pixels lying between some of the H\,{\sc ii}-like agglomerations (e.g., PG\,1011$-$040), but mainly toward the nuclear zones where the H$\alpha$ emission suffers from flux over-subtraction due to AGN deblending. The gas depletion times measured for the star-forming regions within the PG quasar host galaxies are comparable to the average value measured for nearby inactive spirals ($t_{\rm dep} \approx 1.8$\,Gyr; \citealt{Leroy2013}) and the EDGE-CALIFA ($t_{\rm dep} \approx 1.6$\,Gyr; \citealt{Bolatto2017})\footnote{Estimate corrected for the $\alpha_{\rm CO}$ and IMF conventions adopted in this work.} local inactive galaxies at similar kpc scales. Note that the later survey reasonably matches our PG quasar host galaxy sample in terms of molecular gas fraction, SFR, and stellar mass \citep{Molina2021}.

The restframe 230\,GHz continuum data trace zones with higher $\Sigma_{\rm SFR}$ and $\Sigma_{\rm H_2}$, regions where $t_{\rm dep} \approx 0.2-2\,$Gyr.  Those values are in rough agreement with the average estimates computed over the central host galaxy zone (plus signs) where the H$\alpha$ emission is missed due to the inaccurate quasar emission deblending. In PG\,0923+129, the continuum and H$\alpha$ data overlap in terms of $\Sigma_{\rm SFR}$, but this is not the case for the other host galaxies, as the ALMA observations were not sensitive enough to yield reliable detections of the continuum far from the central region. Among our sample of quasar hosts, PG\,0050+124 has the lowest average $t_{\rm dep}$ ($\sim 1.8$\,Gyr), comparable to that seen in normal star-forming disk galaxies. However, the exceptionally high molecular gas surface density of this system leads us to suspect that its gas depletion time may be overestimated. With $\Sigma_{\rm H_2} \approx 1000\,M_\odot$\,pc$^{-2}$, this host galaxy possesses the most compact and densest molecular gas distribution in our sample, suggesting that a lower (``ULIRG-like'') CO-to-H$_2$ conversion factor may be more appropriate for estimating the molecular gas mass \citep{Bolatto2013}.  If so, $\Sigma_{\rm H_2}$ would be lower by a factor of $\sim 3$, and $t_{\rm dep} \approx 0.6$\,Gyr.\footnote{Indeed, detailed analysis of the cold molecular gas dynamics favors a ULIRG-like $\alpha_{\rm CO}$ value for this system (Q. Fei et al., in preparation).} The use of a less conservative estimate of the degree of thermal free-free contamination (e.g., by considering the absorption of Lyman continuum by dust; \citealt{Yun2002}) would systematically increase the dust-based values of $\Sigma_{\rm SFR}$, and hence reduce $t_{\rm dep}$, by 7\,\%--30\,\% for all the host galaxies.

\section{Discussion}
\label{sec:dis}

\begin{figure*}
\centering
\includegraphics[width=2.1\columnwidth]{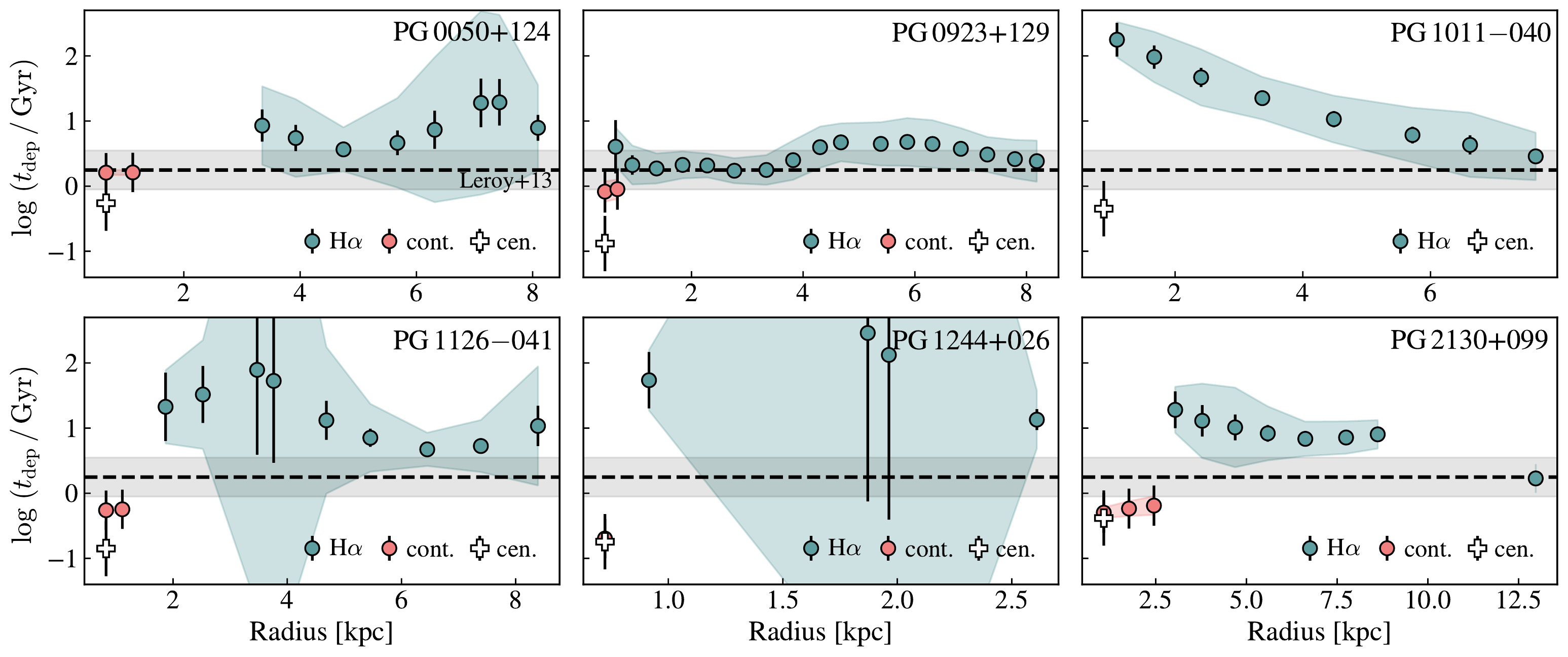}
\caption{\label{fig:tdep_profiles} Radial profiles of depletion time for the six PG quasar host galaxies. The green circles represent the H$\alpha$-based estimates, while the red circles show the estimates using the AGN-decontaminated restframe 230\,GHz continuum data. We also show the average host galaxy central $t_{\rm dep}$ computed over the host galaxy area where we miss the optical-light emission due to AGN deblending flux oversubtraction. The shaded zones indicate the $1\,\sigma$ scatter of the individual pixel data. The dashed line corresponds to the best-fit trend reported for inactive normal galaxies \citep{Leroy2013}.  The radial profiles only account for the data with S/N$\,\geq 3$.}
\end{figure*}

\subsection{Enhanced Star Formation Efficiency}

In apparent contradiction with the expected star formation activity shutoff in AGN host galaxies, the accumulating evidence from recent global measurements attests that the more luminous AGNs are preferentially hosted in gas-rich, highly star-forming galaxies (e.g., \citealt{Bernhard2016,Shangguan2018,Bernhard2019,Shangguan2019,Jarvis2020,Kirkpatrick2020,Xie2021}; but see \citealt{Ward2022}). Indeed, the efficiency of star formation qualifies many host galaxies as starbursts \citep{Shangguan2020b,Koss2021,Zhuang2021}. What is responsible for the apparent coeval episodes of vigorous BH growth and efficient star formation? Tentative evidence suggests that the star formation predominantly occurs on relatively small ($\lesssim 1$\,kpc) central scales \citep{Lutz2016,Zhuang2020}. While major mergers can produce nuclear starburst activity (e.g., local ULIRGs; \citealt{Soifer2001}) and luminous AGNs (e.g., \citealt{Treister2012,Glikman2015}), not all AGN hosts with enhanced star formation activity present dynamical perturbations in their stellar structure \citep{Koss2011,Kim2021,Xie2021} or atomic H\,{\sc i} kinematic perturbations as evidenced from the global line profiles \citep{Ho2008}. The current sample of PG quasars embodies those cases well. Only PG\,0050+124 belongs to a multiple system \citep{Lim1999,Scharwachter2003}, while the rest of the sample display clear, unperturbed disk-like structures as evidenced from broadband HST images analysis \citep{Veilleux2009,Kim2017,Zhao2021}.

The present mapping of the PG quasar host galaxies sheds additional light on properties previously suggested by the unresolved ACA and high-resolution ALMA observations. \citet{Shangguan2020} reported unresolved dust emission in their ACA observations ($\sim 6''$ resolution), suggesting that the mm continuum emission may be predominantly powered by an AGN or a nuclear starburst. \citet{Molina2021} found that the molecular distribution in the PG quasars is compact, with the CO(2--1) half-light radius mainly limited to the $\sim$kpc scale, indicating high molecular gas surface densities and presumably elevated star formation activity. The current study reveals that the ongoing star formation is ubiquitous within the host galaxies, and that their SFR steeply increases toward the central zone, where $\sim 15\,\%-80$\,\% of the total star formation activity budget resides.  The central rise in SFR is driven not merely by the existence of a large molecular gas fuel supply, but also by an elevation of SFE. As Figure~\ref{fig:tdep_profiles} illustrates, the radial profiles of gas depletion times reach $t_{\rm dep} \approx 0.2-2$\,Gyr for the central zones.  We stress that the error budget of the central $t_{\rm dep}$ measurements is dominated by the $\alpha_{\rm CO}$ and SFR$_{\rm IR}$ uncertainties, and not by factors correcting for flux density contamination sources (including the dust heated by the AGN) at the mm wavelength range (Section~\ref{sec:SI_SFR}).  Indeed, these central values of $t_{\rm dep}$ should probably be regarded as upper limits, in view of the conservative corrections for the AGN applied to the mm continuum emission before estimating the SFRs (by $\sim 7\%-30\%$), the CO-to-H$_2$ conversion factor employed to derive molecular gas masses (up to a factor of $\sim 3$), and the presence of unresolved continuum sub-components.  The different timescale sensitivity of the H$\alpha$ and the mm emission as SFR tracers is unlikely to produce the observed radial trends. The molecular gas kinematics of the PG quasar host galaxies are largely regular, with the lack of significant perturbations that could arise from a sudden and drastic change of the cold gas properties due to negative AGN feedback \citep{Molina2021}.

The degree of radial variation in $t_{\rm dep}$, while seen in normal, inactive spirals \citep{Leroy2013}, is not common. \citet{Utomo2017} report that only $\sim 25\,\%$ of the inactive galaxies in the EDGE-CALIFA survey present a mild decrease of $t_{\rm dep}$ toward the nuclear region (down to $\sim 0.7$\,Gyr).\footnote{This value corresponds to $t_{\rm dep} = 1$\,Gyr as reported by \citet{Utomo2017}, after adjusting to the conventions adopted in the current work.}  Galaxies with shorter central gas depletion times are usually associated with more compact molecular and stellar spatial distributions (half-mass radius $\sim 1-4$\,kpc; \citealt{Utomo2017}). The boost of central SFR is more clearly correlated with shortening of $t_{\rm dep}$ in starbursts \citep{Ellison2020}. The most striking radial variations in gas depletion times are seen in local LIRGs, which can exhibit $t_{\rm dep} \lesssim 0.1$\,Gyr inside the central kpc, while $t_{\rm dep} \approx 1$\,Gyr on larger scales \citep{SanchezGarcia2022}. \citet{Molina2021} suggested that the compact molecular gas distributions of the PG quasars are reminiscent of those observed in local LIRGs, whose CO half-light radius is typically $\lesssim 0.8$\,kpc \citep{Downes1998,Iono2009,Bellocchi2022}. Local LIRGs and starbursts also present central sub-kpc scale star formation that contributes more than 50\% of their total IR luminosity (e.g., \citealt{Soifer2001}).  The host galaxies of PG quasars share strong similarities to U/LIRGs in terms of the compactness of their star formation activity and radial variation in gas depletion time.

\begin{figure}
\centering
\includegraphics[width=1.0\columnwidth]{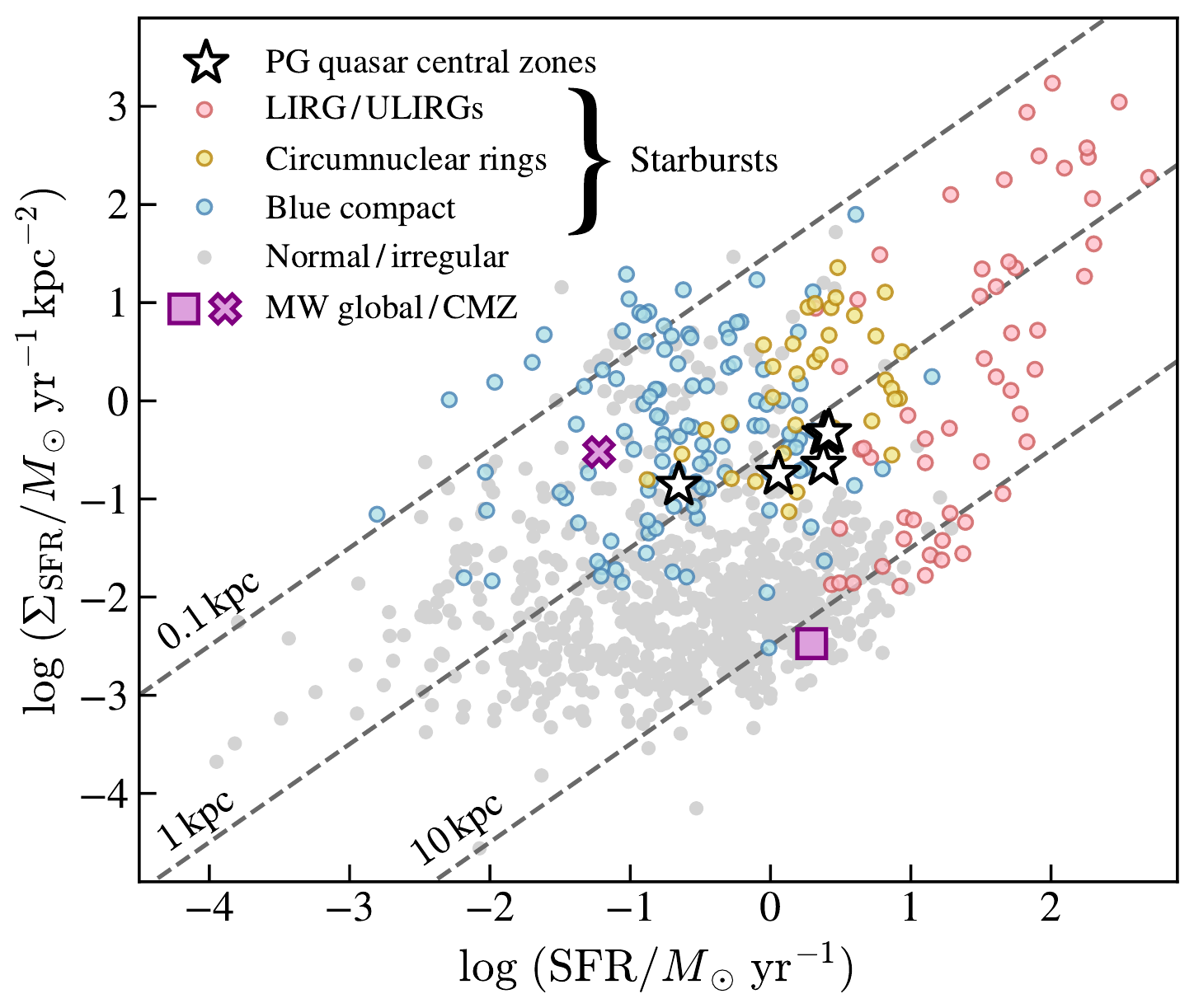}
\caption{\label{fig:SigmaSFR_SFR} The PG quasar host galaxies in the $\Sigma_{\rm SFR}$--SFR plane. We measure both quantities from the central beam-sized zones of the AGN-decontaminated restframe 230\,GHz continuum maps, which implies that the reported $\Sigma_{\rm SFR}$ estimates are lower limits. We compare with the integrated star formation properties of local galaxies and starbursts, including the central molecular zone (CMZ) of the Milky Way (MW) and the circumnuclear rings collated in \citet{Kennicutt2012}. The dashed lines indicate values of constant radius. The central zone of the PG quasar host galaxies display SFRs similar than those measured from local circumnuclear rings and blue compact starburst regions. This figure is adapted from \citet{Kennicutt2012}.}
\end{figure}

Comparison between SFR and $\Sigma_{\rm SFR}$ provides another vantage point to examine the properties of star formation activity in galaxies. \citet{Kennicutt2012} showed that most normal disk galaxies tend to lie in a tight region within this plane, while the starbursts are spread all over, mainly covering the upper 2--3 decades of total SFR and $\Sigma_{\rm SFR}$. Figure~\ref{fig:SigmaSFR_SFR} shows the central PSF-sized zones of the PG quasar host galaxies in the $\Sigma_{\rm SFR}$--SFR plane. While it is not a surprise that the data are clustered around the $\sim 1$\,kpc dashed line because of the limitations imposed by the spatial resolution of the current set of observations, we note that the central regions present $\Sigma_{\rm SFR}$ values high enough to be comparable to those found in circumnuclear rings and blue compact starburst regions. 

Recalling that the mm cold dust continuum is not yet fully resolved, $\Sigma_{\rm SFR}$ in the central regions of the quasars should be even higher than shown. Are we witnessing signs of star-forming nuclear rings in the quasar hosts? At least one-fifth of nearby disk galaxies host star-forming nuclear rings \citep{Knapen2005}, including the Milky Way \citep{Ho1991,Hsieh2017}. Nuclear rings are often reported in observations of less luminous nearby AGNs \citep{Barth1995,Maoz1996,Maoz2001,GarciaBurillo2014,AlonsoHerrero2018,Husemann2019a,Feruglio2020,Winkel2022}, in line with the expectations that the association between star formation and BH accretion seems to occur preferentially on nuclear scales \citep{Davies2007,Watabe2008,Imanishi2011,Canalizo2013,Bessiere2017,Lutz2018,Zhuang2020}. This is not unexpected, as cold gas on small scales naturally fuels nuclear star formation and feeds the BH \citep{HopkinsQuataert2010,Volonteri2015,Gan2019}. Nuclear ring star formation often emerges when pseudo-bulges slowly assemble from disk material \citep{Kormendy2004}. Note that the host galaxies of quasars, especially those classified as ``narrow-line Seyfert~1'', tend to be relatively late-type, often barred, disks with pseudo-bulges \citep{Kim2019,Zhao2021}. Interestingly, \citet{Song2021} measured the amount of the star formation activity taking place in ``nuclear rings'' within local normal galaxies and LIRGs. Defining a nuclear ring as the emitting ring-like structures detected at $3-33$\,GHz within the central 2\,kpc region, they found that $\sim 10\,\%-40\,$\% and $\sim 50\,\%-60\,$\% of the total star formation occurs in the central regions of these normal galaxies and LIRGs, respectively. Those values are consistent with our measurements ($\sim 15\,\%-80$\,\%) based on the mm continuum observations over similar spatial scales.

\subsection{Bulge and Black Hole Growth} 

The correlations between the BH mass and the properties of the galaxy bulges have been commonly interpreted as evidence of a common physical mechanism that drives the BH and galaxy growth \citep{Kormendy2013}. In this section, we focus on the $M_{\rm BH}$--$M_{\rm bulge}$ relation, and we use the molecular gas mass, SFR, BH accretion, and mass outflow rate estimates to investigate how the BH and stellar growth in the host galaxy are synchronized.  We collect bulge half-light radii ($R_e$) and luminosities of the PG quasar host galaxies from \citet{Veilleux2009} and \citet{Zhao2021}, and we convert the bulge luminosities to stellar masses using mass-to-light ratios of \citet{Bell2001}, following \citet{Zhao2021}. The BH mass estimates are presented in Table~\ref{tab:sample}.

\begin{figure}
\centering
\includegraphics[width=1.0\columnwidth]{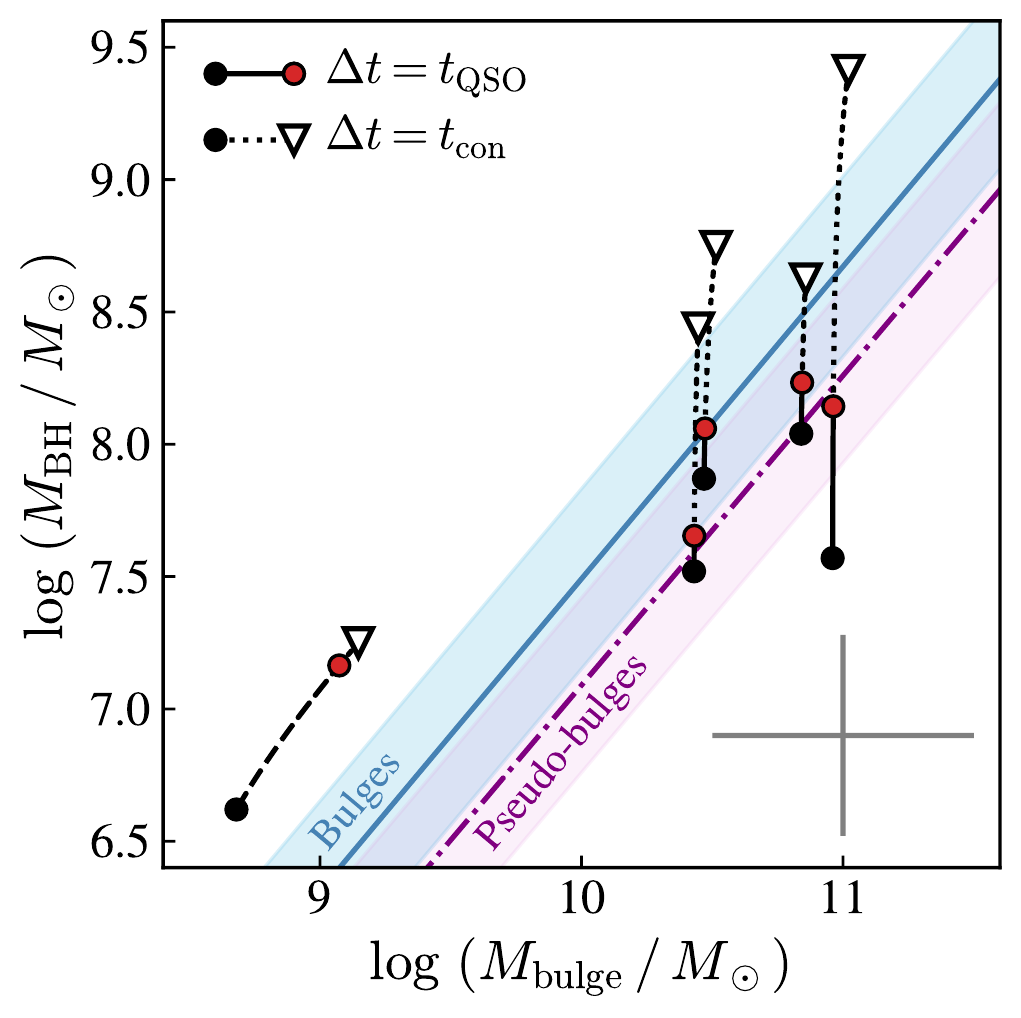}
\caption{\label{fig:Bulge-BH_coev} The PG quasar host galaxies in the BH mass-bulge mass plane. The typical data uncertainty is shown in the bottom-right corner.  The blue solid and the purple dot-dashed lines represent the relations for bulges and pseudo-bulges taken from \citet{Zhu2021} and \citet{Li2022}, respectively. The color-shaded zones represent the scatter ($\sim 0.34$\,dex) for each relation. We present mass evolutionary tracks up to the time in which the molecular gas is consumed ($t_{\rm con}$) within 2 times the bulge half-light radius. The consumption timescale estimates account for the molecular gas mass, SFR, mass outflow rate, and BH accretion rate, assuming no variation in the mass growth rate and a closed-box model. We also highlight the upper limit of the quasar lifetime ($t_{\rm QSO} \approx 0.1\,$Gyr; \citealt{Martini2004}), which in all cases is shorter than $t_{\rm con}$. The dashed line marks PG\,1244+026, which has no available mass outflow rate and is omitted from the mass evolutionary track estimation.}
\end{figure}

In Figure~\ref{fig:Bulge-BH_coev} we show the PG quasars host galaxies in the $M_{\rm BH}$--$M_{\rm bulge}$ plane, along with the relations for bulges (classical bulges and the cores of ellipticals; see \citealt{Zhu2021}) and pseudo-bulges \citep{Li2022}. Four of the six PG quasars are below the standard relation for bulges, but they lie on the relation for pseudo-bulges, consistent with the morphology of their central spheroid reported by \citet{Zhao2021}. PG\,0050+124 (I~Zwicky 1) falls $\sim 1.2$\,dex below the zero point of the relation for pseudo-bulges \citep{Ding2022}, while PG\,1244+026 is the only system that lies above the relation for bulges. Before computing the mass growth tracks of the BH and pseudo-bulge, we note that simple consideration of the CO(2--1) data indicates that, apart from PG\,1244+026, whose molecular gas mass is twice the bulge stellar mass, the molecular gas fractions within $2\,R_e$ of the pseudo-bulges are merely $\sim 0.06-0.21$. This suggests that, even in the unrealistic case that all of the central molecular gas gets converted into stars, the pseudo-bulges will not increase their stellar mass significantly unless more gas is funneled into the nuclear region of the galaxy. Except for PG\,1244+026, the effective radius of the molecular gas distribution is comparable to that of the pseudo-bulge, which implies that there is not enough molecular gas within the host to produce substantial stellar mass growth of the pseudo-bulge. Any new source of gas that may substantially grow the mass of the pseudo-bulge should be external to the host galaxy disk. In the absence of such a gas accretion event, only the BH can increase its mass significantly. 

Can the BHs accrete gas rapidly enough to migrate upward to the locus of the standard $M_{\rm BH}-M_{\rm bulge}$ relation for bulges? To test this, we compute mass growth tracks for the BH and the pseudo-bulge by assuming (1) that the BH accretion rate $\dot{M}_{\rm BH} = L_{\rm bol} / \epsilon \, c^2$, where $L_{\rm bol}$ is the bolometric luminosity of the quasar (Table~1), $\epsilon = 0.1$ is the radiative efficiency for a standard accretion disk around a Schwarzschild BH, and $c$ is the speed of light, and (2) that the pseudo-bulge grows according to the SFR (estimated from the restframe 230\,GHz continuum) and the molecular gas content within $2\,R_e$.  We also consider the mass loss attributed to the ionized gas outflows \citep{Molina2022}, if available,\footnote{We adopt the ionized gas mass outflow rates computed by assuming an expanding shell-like shock front outflow geometry (see \citealt{Molina2022}, for more details).} but we do not account for molecular outflows because of the lack of robust detection \citep{Shangguan2020b}. Assuming a closed-box model of constant SFR and mass outflow rate, we show in Figure~\ref{fig:Bulge-BH_coev} the mass growth tracks for a molecular gas consumption timescale of $t_{\rm con} \approx 0.1-2$\,Gyr. We omit PG\,1011$-$040 because it lacks a central SFR from dust emission. The mass growth tracks suggest that the BHs could significantly increase their mass, in contrast to the stellar mass of the pseudo-bulge, which, as mentioned before, will hardly change\footnote{ The exception is PG\,1244+026, whose bulge can significantly increase its stellar mass, making the mass growth track parallel to the standard $M_{\rm BH}-M_{\rm bulge}$ relation.}. The BH accretion in principle is high enough to elevate the BH masses to the locus of inactive classical bulges and ellipticals.  If this were true, it would present a problem in view of the fact that these sources appear to be hosted in pseudo-bulges \citep{Zhao2021}. This difficulty can be reconciled by recognizing that the expected quasar lifetime, $t_{\rm QSO} \approx 0.1\,$Gyr (\citealt{Martini2004}), is shorter than $t_{\rm con}$.

\section{Conclusions}
\label{sec:con}

We combine ALMA and MUSE observations of six nearby $z \lesssim 0.06$ Palomar-Green quasar host galaxies, delivering detailed maps of the molecular gas, mm continuum, and ionized gas emission at kpc scales. We carefully remove the contribution of the AGN to the IR continuum and optical emission-line flux to yield spatial distributions of SFR. Through detailed comparison with the molecular gas data, we further compute SFEs to investigate the nature of the star formation activity within the quasar hosts, drawing comparisons with the general local galaxy population. The data suggest that the global far-IR SFRs and SFEs reported for the quasar host galaxies mainly reflect the physical conditions that drive the ongoing star formation activity in their nuclear, kpc-scale zone. These central zones present enhanced SFEs and abnormally high SFR surface densities, estimates that are consistent with those reported for nearby nuclear starbursts. We summarize our main conclusions as follows:

\begin{itemize}

\item The mm continuum emission at restframe 230\,GHz, detected in the central $\lesssim 2$\,kpc region of the host galaxies, mainly corresponds to the dust emission produced by obscured star formation activity, with minor, but not negligible ($\lesssim 40$\,\%), contribution from synchrotron and thermal free-free emission. The central zones of the PG quasars form stars at a rate of $0.5-7\,M_\odot$\,yr$^{-1}$, encompassing $\sim 15\,\%-80$\,\% of the total SFRs derived from the global far-IR SED measurements. 

\item The H$\alpha$ emission traces ongoing star formation activity across the host galaxy, although our ability to properly recover the line flux near the quasar ($\lesssim 2-3$\,kpc) is frustrated by inaccurate spectral deblending. Nevertheless, at large galactocentric radius, we find that the star-forming regions are mainly located within spiral arms, inner rings, and H{\sc ii}-region complexes. Those regions account for total SFR$_{\rm H\alpha} \approx 0.3-3.8\,M_\odot$\,yr$^{-1}$ after correcting for dust attenuation and AGN contamination.

\item In comparison to the molecular gas distribution, the SFR rises more steeply toward the nucleus of the host galaxies, which implies that the central zone has enhanced SFEs or shortened gas depletion times. With $t_{\rm dep} \approx 0.2-2$\,Gyr, the central zones can be deemed starburst regions. By contrast, in the outskirts of the host galaxy the star formation activity ensues at efficiency levels comparable to that reported for normal, star-forming galaxies ($t_{\rm dep} \approx 1.8$\,Gyr).

\item The observed central SFRs and molecular gas will hardly lead to a significant increase of the stellar mass of the bulge. The BH mass is expected to undergo modest gains ($\sim 0.3\,$dex) while still obeying the $M_{\rm BH}-M_{\rm bulge}$ relation of inactive galaxies.
\end{itemize}

\acknowledgments{We thank to the referee for helpful suggestions.  We acknowledge support from the National Science Foundation of China (11721303, 11991052, 12011540375) and the China Manned Space Project (CMS-CSST-2021-A04, CMS-CSST-2021-A06).  R. W.~acknowledges support from the National Science Foundation of China (NSFC) grant No. 12173002. This work was funded by ANID - Millennium Science Initiative Program - ICN12\_009 (F.E.B.), n\'ucleo Milenio TITANs - NCN198\_058 (E.T.),  CATA-BASAL - ACE210002 (F.E.B. and E.T.) and FB210003 (F.E.B. and E.T.), and FONDECYT Regular - 1190818 (F.E.B. and E.T.) and 1200495 (F.E.B. and E.T.). This paper makes use of the following ALMA data: ADS/JAO.ALMA\#2018.1.00006.S. ALMA is a partnership of ESO (representing its member states), NSF (USA) and NINS (Japan), together with NRC (Canada), MOST and ASIAA (Taiwan), and KASI (Republic of Korea), in cooperation with the Republic of Chile. The Joint ALMA Observatory is operated by ESO, AUI/NRAO and NAOJ.  This research has made use of the services of the ESO Science Archive Facility, and based on observations collected at the European Organization for Astronomical Research in the Southern Hemisphere under ESO programme IDs 094.B$-$0345(A), 095.B$-$0015(A),  0103.B$-$0496(B),  and 0104.B$-$0151(A).}

\software{\textsc{Astropy}\,\citep{astropy:2013,astropy:2018}, \textsc{lmfit}\,\citep{Newville2014}, \textsc{matplotlib}\,\citep{Hunter2007}, \textsc{numpy}\,\citep{Oliphant2006}, \textsc{photutils}\,\citep{Bradley2020}, \textsc{scikit-image}\,\citep{scikit-image}, \textsc{scipy}\,\citep{scipy2020}.}
\bigskip
\bigskip
\bigskip
\appendix

\section{Far-IR and Radio Continuum Measurements}
\label{sec:AppA}

Table~\ref{tab:radio_data} summarizes the far-IR and radio continuum measurements from the literature used to construct the broad-band SEDs analyzed in Figure~\ref{fig:submmSED} and Section~3.3. 

\begin{table*}
	\centering
	\def\arraystretch{1.0}
	\setlength\tabcolsep{3pt}
    	\caption{\label{tab:radio_data} Far-IR and Radio Continuum Measurements}
    	\vspace{0.2mm}
	\begin{tabular}{cccc}
		\hline
		\hline
		Object & Frequency & $S_{\nu}$ & Reference \\
		& (GHz) & (mJy) & \\
		(1) & (2) & (3) & (4) \\
		\hline
	PG\,0050+124 & 3000 & $2959\pm51$ & \citet{Hughes1993} \\
	& 666.2 & $225\pm8$ & \citet{Hughes1993} \\
	& 374.7 & $18.4\pm4.4$ & \citet{Hughes1993} \\
	& 45.0 & $0.30\pm0.08$ & \citet{Baldi2022}$^a$ \\
	& 14.9 & $1.06\pm0.32$ & \citet{Barvainis1996} \\
	& 8.43 & $1.02\pm0.03$ & \citet{Yang2020} \\
	& 5.00 & $2.60\pm0.07$ & \citet{Kellermann1994} \\
	& 4.90 & $3.10\pm0.20$ & \citet{Edelson1987} \\
	& 4.90 & $1.80\pm0.11$ & \citet{Barvainis1996} \\
	& 4.89 & $2.21\pm0.19$ & \citet{Barvainis1996} \\
	& 4.86 & $1.85\pm0.11$ & \citet{Yang2020} \\
	& 3.00 & $4.05\pm0.27$ & \citet{Gordon2021}$^b$ \\
	& 1.49 & $6.22\pm0.56$ & \citet{Barvainis1996} \\
	& 1.42 & $4.91\pm0.27$ & \citet{Yang2020} \\
	& 1.40 & $8.40\pm0.90$ & \citet{Edelson1987} \\
	& 1.40 & $8.80\pm0.60$ & \citet{Condon1998} \\
	& 0.685 & $11.0\pm1.0$ & \citet{Silpa2020} \\
         PG\,0923+129 & 8.46 & $2.0\pm0.05$ & \citet{Schmitt2001} \\
         & 5.00 & $3.31\pm0.04$ & \citet{Berton2018} \\
         & 5.00 & $10.0$ & \citet{Kellermann1994} \\
         & 3.00 & $5.00\pm0.25$ & \citet{Gordon2021}$^b$ \\
         & 1.40 & 9.00 & \citet{VeronCetty2010} \\
         & 1.40 & $8.53\pm0.14$ & \citet{Helfand2015} \\
         & 1.40 & $7.45\pm0.15$ & \citet{Helfand2015} \\
         & 0.685 & $14.0\pm1.0$ & \citet{Silpa2020} \\
         PG\,1011$-$040 & 5.50 & $0.38\pm0.02$ & \citet{Chen2020radio} \\
         & 5.00 & 0.28 & \citet{Kellermann1994} \\
         & 4.86 & $0.48\pm0.11$ & \citet{Yang2020} \\
         & 0.685 & $1.6\pm0.1$ & \citet{Silpa2020} \\
         PG\,1126$-$041 & 5.00 & 0.51 & \citet{Kellermann1994} \\
         & 3.00 & $1.17\pm0.35$ & \citet{Gordon2021}$^b$ \\
         & 1.40 & 1.23 & \citet{Helfand2015} \\
         & 1.40 & 1.11 & \citet{Helfand2015} \\
         PG\,1244+026 & 5.17 & $0.85\pm0.03$ & \citet{Yang2020} \\
         & 5.00 & 0.83 & \citet{Kellermann1994} \\
         & 5.00 & $0.70\pm0.03$ & \citet{Berton2018} \\
         & 3.00 & $2.00\pm0.51$ & \citet{Gordon2021}$^b$ \\
         & 1.40 & $2.23\pm0.14$ & \citet{Rafter2009}\\
         & 0.685 & $3.10\pm0.3$ & \citet{Silpa2020} \\
         PG\,2130+099 & 14.9 & $1.47\pm0.35$ & \citet{Barvainis1996} \\
         & 8.60 & $1.50\pm0.15$ & \citet{Barvainis2005} \\
         & 5.00 & 2.05 & \citet{Kellermann1994} \\
         & 4.89 & $2.09\pm0.18$ & \citet{Barvainis1996} \\
         & 3.00 & $3.10\pm0.32$ & \citet{Gordon2021}$^b$ \\
         & 1.49 & $4.71\pm0.48$ & \citet{Barvainis1996} \\
         & 1.40 & $6.50\pm0.50$ & \citet{Condon1998} \\
         & 0.685 & $8.70\pm0.70$ & \citet{Silpa2020} \\         
		\hline
	\end{tabular}
	\justify
	{\justify \textsc{Note}--- (1) Source name. (2) Restframe frequency. (3) Flux density. (4) Reference. The rest of the far-IR SED estimates are reported in \citet{Shangguan2018}. $^a$ Peak flux density value in units of mJy\,beam$^{-1}$, for a marginally resolved detection.$^b$ The flux density estimates were corrected for $\sim 9$\,\% systematic underestimation \citep{Gordon2021}.}
\end{table*}

\section{Systematic Uncertainties of the Mixing-sequence Method}
\label{sec:AppC}

\begin{figure}
\centering
\figurenum{B1}
\includegraphics[width=0.9\columnwidth]{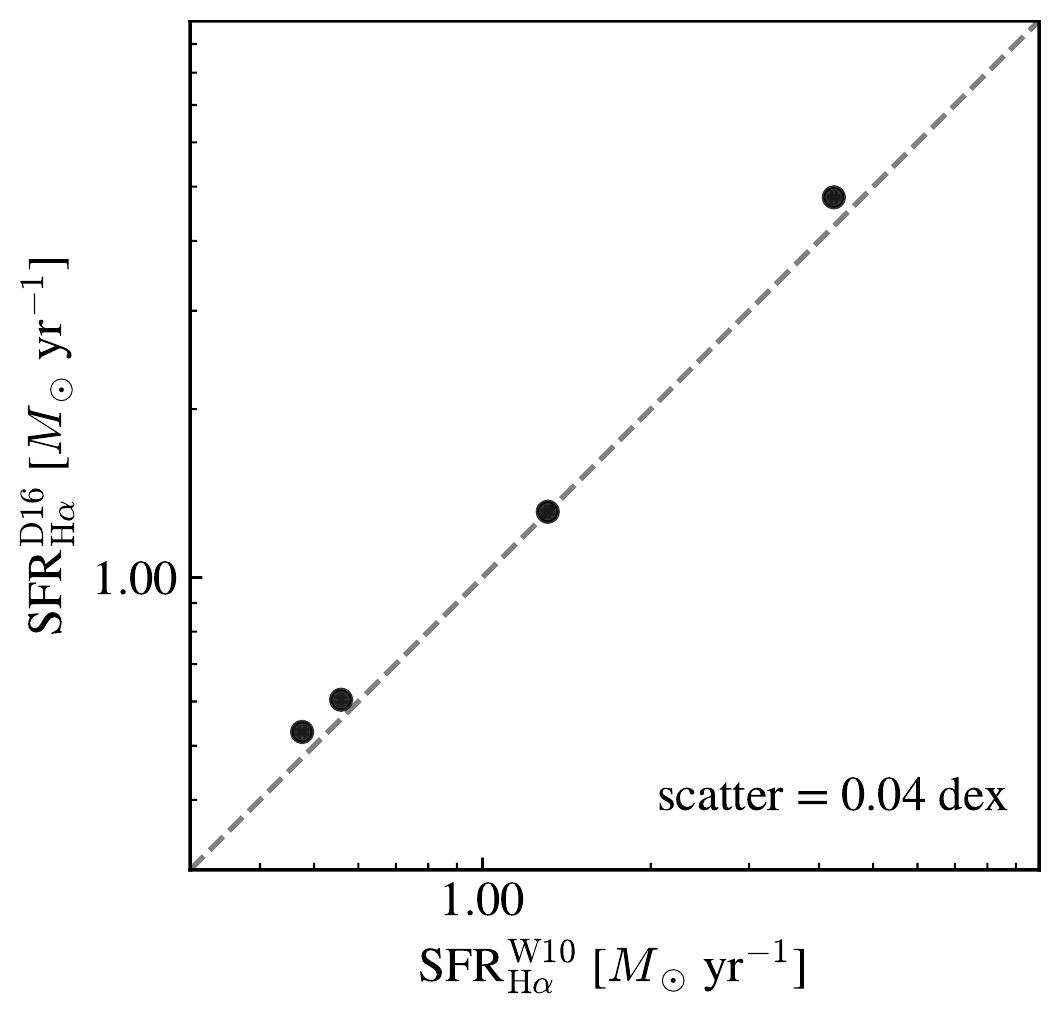}\\
\caption{\label{fig:SFR_mixing_comp} Comparison between the global SFR$_{\rm H\alpha}$ estimated by applying the two mixing-sequence corrections adopted in this work.  We only provide  SFR$^{\rm D16}_{\rm H\alpha}$ estimates for four host galaxies due to method limitation}. The dashed line represents the 1:1 ratio.  The small scatter suggests that the global SFR$_{\rm H\alpha}$ estimates are not particularly sensitive to the adopted mixing-sequence method.
\end{figure}

\begin{figure}
\centering
\figurenum{B2}
\includegraphics[width=0.9\columnwidth]{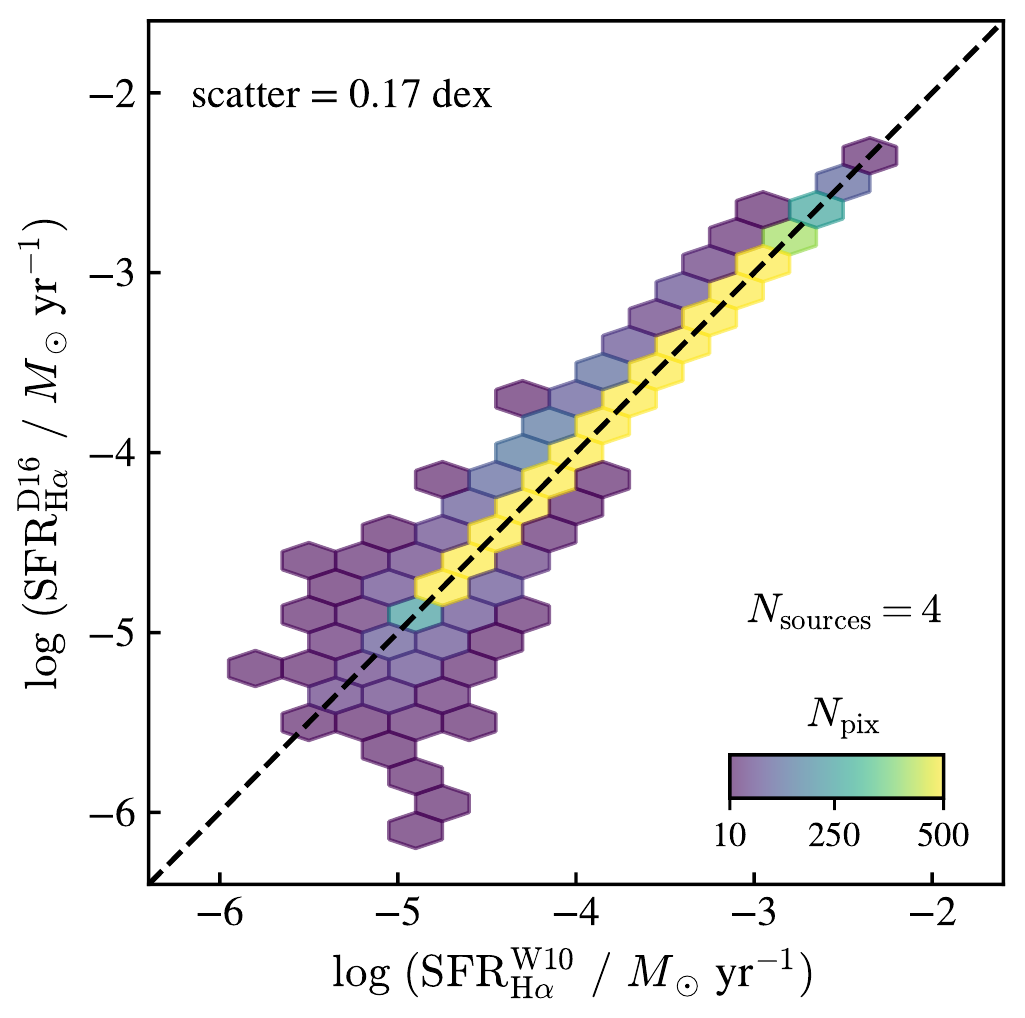}\\
\caption{\label{fig:pw_SFR_mixing_comp} Pixel-wise comparison between the SFR values estimated by applying the two mixing-sequence corrections adopted in this work. The dashed line represents the 1:1 ratio.  We only report SFR$^{\rm D16}_{\rm H\alpha}$ for the four host galaxies with available estimates.  The data far from the 1:1 ratio tend to comprise fewer pixels and are associated mainly with regions with fainter H$\alpha$ emission.}
\end{figure}

The use of different mixing-sequence correction methods to compute the SFR in AGN host galaxies naturally leads to estimates that differ from each other. Figure~\ref{fig:SFR_mixing_comp} compares global SFR$_{\rm H\alpha}$ values derived after applying the two mixing-sequence correction methods adopted in this work. We find that the global SFR$_{\rm H\alpha}$ estimates are not particularly sensitive to the applied mixing-sequence correction, with the discrepancies limited to $\lesssim 10\%$. In Figure~\ref{fig:pw_SFR_mixing_comp} we show the pixel-wise comparison of the SFR. The scatter increases for regions with low SFR estimates, although the pixels far from the 1:1 line are few and individually contribute little to the total SFR of each galaxy. The deviant pixels trace regions with elevated [O\,{\sc iii}]/H$\beta$, ones that lie close to or above the maximum starburst curve \citep{Kewley2001} in the [N\,{\sc ii}]/H$\alpha$--[O\,{\sc iii}]/H$\beta$ diagram.

\section{Notes for Individual Galaxies}
\label{sec:AppD}

\begin{itemize}

\item[]\textbf{PG\,0050+124:} The restframe 230\,GHz continuum is barely resolved following the visibility amplitude profile in the $u-v$ plane and the \texttt{uvmodelfit} best-fit model. The continuum emission is consistent with that predicted from the global far-IR SED fitting \citep{Shangguan2018}, suggesting that this compact emission is mainly tracing the obscured star formation activity occurring in the central region of this host galaxy. 

\item[]\textbf{PG\,0923+129:} The continuum emission is extended, and can described by a point-like source sub-component plus an exponential profile. The point-like source sub-component is consistent with being powered by the synchrotron emission following the extrapolation from the radio SED.  A tail-like feature is seen toward the west at the $2\,\sigma$ level in the two-dimensional map of the continuum.`

\item[]\textbf{PG\,1011$-$040:} The restframe 230\,GHz continuum is marginally detected at the $3\,\sigma$ level. The visibility data indicate that this emission is unresolved.

\item[]\textbf{PG\,1126$-$041:} A clear elongated continuum component across the major axis of the host galaxy can be seen. The data are well-described by a point-like source sub-component plus a Gaussian profile. From the SED analysis, most of the continuum emission is powered by the obscured star formation activity, including the point-like source sub-component; however, the radio SED is poorly constrained for this source.

\item[]\textbf{PG\,1244+026:} From the ALMA observation we see an extended asymmetric and bright (S/N\,$>5$) continuum sub-component toward the southwest direction. The asymmetric sub-component is not detected in CO(2--1) emission nor in H$\alpha$ or other optical emission lines, but it is co-spatial with a low-S/N radio emission detection at 0.685\,MHz \citep{Silpa2020}. The restframe 230\,GHz continuum is overly bright with respect to the SED model prediction of \citet{Shangguan2018}, but once the central compact emission is isolated, we measure a flux density consistent with that predicted by the SED model. We speculate that the asymmetric sub-component may be a dusty and diffuse cold-gas outflow not traced (or completely outshined) in the Herschel bands. The presence of an obscured second nucleus is also a possibility; however, the lack of a companion detected in ionized gas or stellar light makes this possibility unlikely.

\item[]\textbf{PG\,2130+099:} The continuum emission is resolved and well-described by a point-like source plus an additional Gaussian component. Two minor tails can be seen at the $2\,\sigma$ level in the east-west orientation. The central point-like source is consistent with cold dust emission associated with underlying star formation activity.

\end{itemize}

\bibliography{bibliography}
\bibliographystyle{aasjournal}
\end{document}